\documentclass[traditabstract]{aa}  
\usepackage{graphicx}
\usepackage{longtable}
\usepackage{natbib}
\usepackage{amsmath}            
\usepackage{txfonts}

\newcommand{\ratioo} {N({\rm H}_2) / I_{\rm CO}}

\newcommand{\kms}   {{\rm \  km \  s^{-1}}}
\newcommand{\K}   {{\rm \  K}}

\newcommand{\lsun}{L$_{\odot}$}   
\newcommand{\msun}{M$_{\odot}$}

\begin{document}

\title{ Luminosity functions and IMF variations from large samples of HII regions and molecular clouds }
                
\author{J. Braine\inst{1}, E. Corbelli\inst{2}  }  

\institute{Laboratoire d'Astrophysique de Bordeaux, Univ. Bordeaux, CNRS, B18N, all\'ee Geoffroy Saint-Hilaire, 33615 Pessac, France \\
             \email{jonathan.braine@u-bordeaux.fr}
\and   INAF-Osservatorio Astrofisico di Arcetri, Largo E. Fermi, 5, 50125 Firenze, Italy \\
\email{edvige.corbelli@inaf.it}
}

\abstract {Large high-quality samples of HII regions and their parent Giant Molecular Clouds (GMC) are now available for local galaxies.  
It is therefore possible to investigate links between the CO and H$\alpha$ luminosity functions and whether massive stars form 
in GMCs of all masses. 
The CO luminosity functions (LF), representing the distribution of GMC masses, are consistently steeper than the H$\alpha$ luminosity functions.  The CO LF invariably steepens in the outer disk where fewer massive GMCs are present beyond the median cloud galactocentric distance.  The H$\alpha$ LF also steepens in the outer disk for most of the galaxies examined.  
Using Salpeter, Kroupa, and Chabrier Initial Mass Functions (IMF) along with stellar mass-luminosity-radius relations, we compute numerically the 
bolometric luminosity and H$\alpha$ emission from young star clusters. The cluster masses are linked to the GMC mass by assuming that the cluster mass is a constant fraction (3\%) of the parent cloud mass.  In particular, results for a fully stochastic IMF are compared to suggestions
that very massive stars only form in massive clusters or clouds.  Within the limits of the observations -- no small molecular clouds or low-luminosity HII regions can be detected at the typical $\sim 10$~Mpc distance of the sample galaxies -- we find no evidence for a maximum stellar mass which varies with cloud or cluster mass.}


\maketitle
\nolinenumbers
 
\section{Introduction}
What can the 
large samples of HII regions and Giant Molecular Clouds (GMCs) now available tell us about the stellar Initial Mass Function (IMF)? 
In particular, is the IMF the same for large and small GMCs?
It is commonly accepted that stars, and thus HII regions, form in GMCs.  
If small GMCs do not form massive stars,
i.e. they do not populate the upper end of the stellar mass distribution, 
then this should be visible in the HII region luminosity function.
This is the main subject of this short contribution.

As in previous works, we will assume that the intrinsic shape of the GMC mass distribution (or function) is that of a power law $N(m)dm \propto m^{-\alpha} dm$, with or without truncation at the high mass end
\citep{Solomon87,Rosolowsky07,Gratier12,Colombo2014a,Braine2018,Rosolowsky2021}.
When the index of the mass (or luminosity) 
function is steeper than $\alpha=2$, then most of the mass is in the small entities.
The lifetimes of the massive stars creating HII regions are quite short so large samples are required to sample all phases (ages) of young stellar clusters (YSCs).

While the stellar initial mass function (IMF) seems fairly constant in today's spiral disks, it appears  growingly accepted that the IMF in low-metallicity or extremely dense environments is different \citep{Larson1998,Kroupa2001,Hopkins2018}.  The change is likely a shift in the characteristic (i.e. average) mass $M_{char}$ of  low-mass stars, without changing the formation process of massive stars and hence leaving the slope of the high-mass part of the IMF unchanged \citep{Hopkins2018}.  This would be due to an increase in core Jeans mass due to a lower metal (dust) content or to factors inhibiting fragmentation (e.g. tidal forces).  Increasing $M_{char}$ means that a smaller fraction of the gas converted into stars goes into low-mass stars, raising the luminosity-to-mass ratio (L/M) of the cluster.
 
In this work we compare the GMC luminosity function
with the HII region luminosity function in local spiral galaxies.  They are traced respectively by the CO(2--1) and H$\alpha$ lines.  The CO luminosity is generally used as an equivalent of cloud mass.
Since the range in HII region or GMC luminosity 
between completeness and a possible truncation at the high mass end is often small, considerable attention will be paid to the fitting.

The lifetime of a GMC $t_{\rm GMC}$, that is to say the time during which it is recognizable as a GMC, is of order 15 Myr \citep{Gratier12,Corbelli17,Calzetti2012,Chevance2020,Kobayashi2017,Demachi2024}.
At a galactic scale, the H$_2$ depletion time $t_{\rm depl} = M(H_2)/SFR$ where SFR is the star formation rate in solar masses per year, 
is about $2 \times 10^9$ yrs \citep{Murgia02,Leroy2013}.  At a galactic scale, $SFR = \epsilon_{SF} M(H_2) / t_{\rm GMC}$ so, 
averaged over the whole star formation cycle, the fraction of the gas turned 
into stars is about $\epsilon_{SF} = t_{\rm GMC}/t_{\rm depl} \approx 1\%$.   Within the gas sufficiently dense to be identified by cloud-finding 
algorithms (typically a few times the rms noise), the efficiency $\epsilon_{SF}$ appears to be higher, some 2-3\% \citep{Murray2011,Evans2009}, 
and the difference can be explained by the fact that of order half of the gas detected at galactic scales is not locked into individual clouds 
because it is too diffuse and does not emit strongly enough in the CO lines.

\citet{Larson2003} found that the mass of the most massive star of a cluster increased with the mass of the star cluster roughly as 
$M_{*,max} \propto M_{clust}^{0.45}$.  
For a given $\epsilon_{SF}$, this suggests that the stellar mass range in a cluster depends on the GMC mass and indeed the original 
study by \citet{Larson1982} found $M_{*,max} = 0.33 M_{\rm GMC}^{0.43}$. 
For a 10000~\msun\ cloud, the IMF would be truncated at $\sim20$~\msun\ in this scenario. 
A similar scenario, also limiting the production of high-mass stars in small clusters, was proposed by \citet{Weidner2006}.  As a result of either of these scenarii, a galaxy would have fewer massive stars than predicted by the IMF because low-mass GMCs would not contribute to the high-mass star population. On the other hand, if the IMF is stochastically sampled 
then although the birth of a massive star is less likely in a low mass cluster, they occasionally form and the overall IMF 
is preserved \citep{2009A&A...495..479C,daSilva2014}.
The data examined here enable us to examine possible links between cloud mass and the mass of the most massive star formed because the H$\alpha$ luminosity of a cluster, and particularly a small cluster, 
depends strongly on the mass of the most massive star remaining.

Large homogeneous high-quality samples of H$\alpha$ and CO observations of individual clouds and HII regions in nearby galaxies have only recently become available.
 The PHANGS \citep[Physics at High Angular resolution in Nearby GalaxieS][]{Leroy2021} survey has made public 
CO and H$\alpha$ luminosities for large samples of GMCs \citep{Rosolowsky2021} and HII regions \citep{Santoro2022} in the  galaxies of their sample.
 As the overlap of the CO and H$\alpha$ samples is small (2 galaxies), we add data for the Local Group galaxy M33 as outlined below in Section 2.  Section 3 describes how we link cloud masses to young stellar cluster masses and to populations of individual stars and their properties, along with the fitting methods. In Section~4 we compare the results of the 
fits to the HII region and cloud luminosity functions for the inner and outer parts of the galactic disks.  The distributions of GMC luminosities are systematically steeper than those of HII regions and Section 5 examines possible explanations and particularly whether introducing a stellar mass limit \citep{Larson1982,Weidner2006} helps explain  the results.

\section{Samples}

We selected galaxies with enough data to be suitable for statistical analyses.  On average, the GMC and HII region samples  have about 700 GMCs and 1000 HII regions per galaxy.  They are described below.  With the exception of M33, the resolutions of the CO and H$\alpha$ observations are similar (see Table 2 of \citet{Rosolowsky2021} and Table 1 of \citet{Santoro2022}). However, in all cases, we use the catalogs produced from the observations by the observers.

\subsection {Molecular clouds}

As part of the PHANGS collaboration \citep[e.g.][]{Schinnerer2019}, a large catalogue of GMCs observed with ALMA was published 
and made public by \citet{Rosolowsky2021}.  They present data for 10 galaxies 
but we do not include data for two small galaxies 
with too few clouds (less than 100) to determine a mass function.  The 8 remaining galaxies are large spirals with effective radii 
between 2.6 and 4.1 kpc and have between 275 and 1432 clouds (over 5000 for the 8 galaxies, see their Table 5).

We add data for the local group spiral M33 \citep{Druard14,Corbelli17} for which both CO and H$\alpha$ data are available.  
Basic information on the galaxies can be found in Table 1 of \citet{Rosolowsky2021} and in Table 1 of \citet{Druard14} for M33.

\subsection{HII regions}

 Also as part of PHANGS, a large ($>20000$) catalogue of HII regions observed with MUSE was published and made publicly 
 available by \citet{Santoro2022}.  19 galaxies were observed with 472 to 2536 HII regions identified per galaxy.  Basic information on the galaxies can be found in Tables 1 and 2 of \citet{Santoro2022}.  
The distances and angular and physical resolutions of the observations are similar to those of \citet{Rosolowsky2021} but, despite being 
part of the PHANGS sample as well, only two galaxies were observed in common.
Adding M33 H$\alpha$ data from \citet{Lin2017} provides a third galaxy in common.

Fitting power laws to the distribution of H$\alpha$ luminosities  for each galaxy, \citet{Santoro2022} find that the spectral index $\alpha$ may decrease (flatten) with 
increasing galaxy-averaged star formation rate surface density.  The range in $\alpha$ was from 1.52 to 2.04 with formal uncertainties of about 0.1.  
Identifying spiral and interarm regions, they find that $\alpha$ is steeper in the interarm.
They did not find a change in $\alpha$ between the inner and outer parts of the galaxies, unlike what was found for GMCs by 
\citet{Gratier12} and \citet{Rosolowsky05}, raising the question of why would (or how could) the HII region and GMC distributions be different.

Finally, we noticed that the spectral index $\alpha$ depended on the minimum luminosity deduced by the powerlaw.py algorithm 
they used \citep{Alstott2014}, pushing us to investigate further thsi issue (see below).

\begin{figure}
        \centering
        \includegraphics[width=\hsize{}]{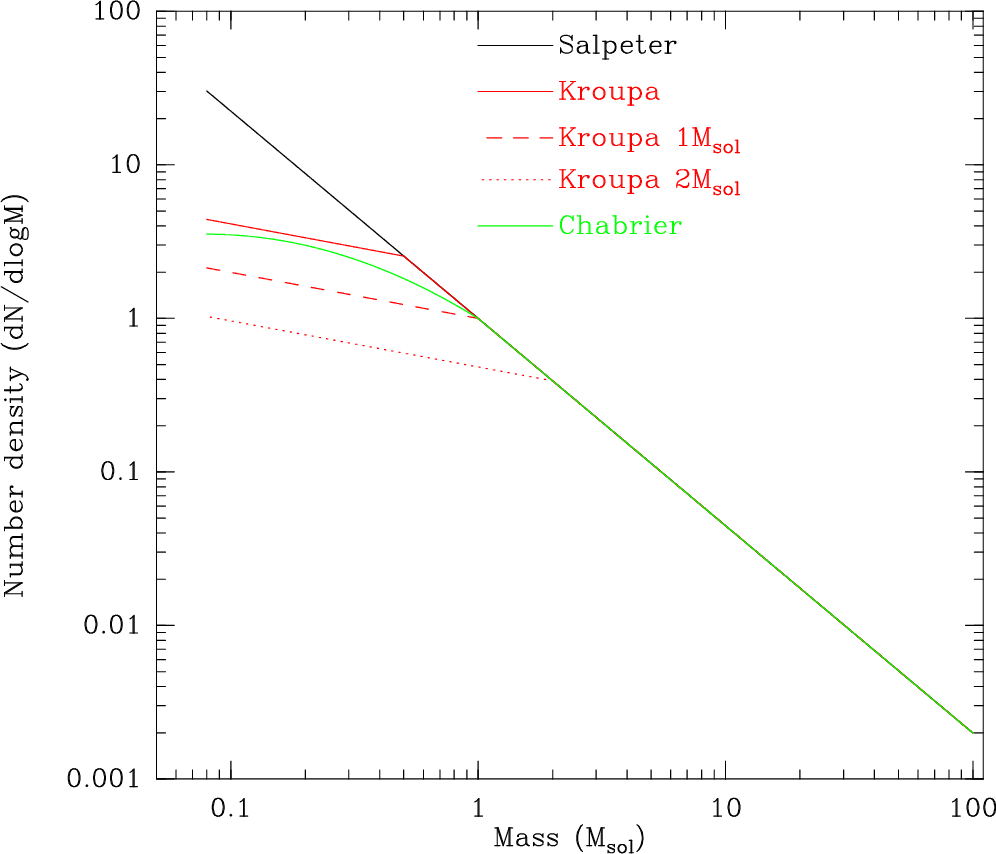}
        \caption{Initial Mass Functions used in this work. Black line is Salpeter, red is Kroupa, and green the Chabrier IMF.  The plot shows the number 
        of stars per logarithmic mass interval for each mass. Clearly, the Kroupa and Chabrier IMFs are similar 
        and have significantly fewer low-mass stars than the Salpeter IMF.  The dashed and dotted lines show the Kroupa IMF with somewhat higher characteristic masses. The effect of increasing the characteristic mass is that there are fewer low mass stars so the L/M ratio of the stellar population increases from about 900 \lsun/\msun to 1500 \lsun/\msun. 
        }
        \label{imf} 
\end{figure}

\begin{figure}
        \centering
        \includegraphics[width=\hsize{}]{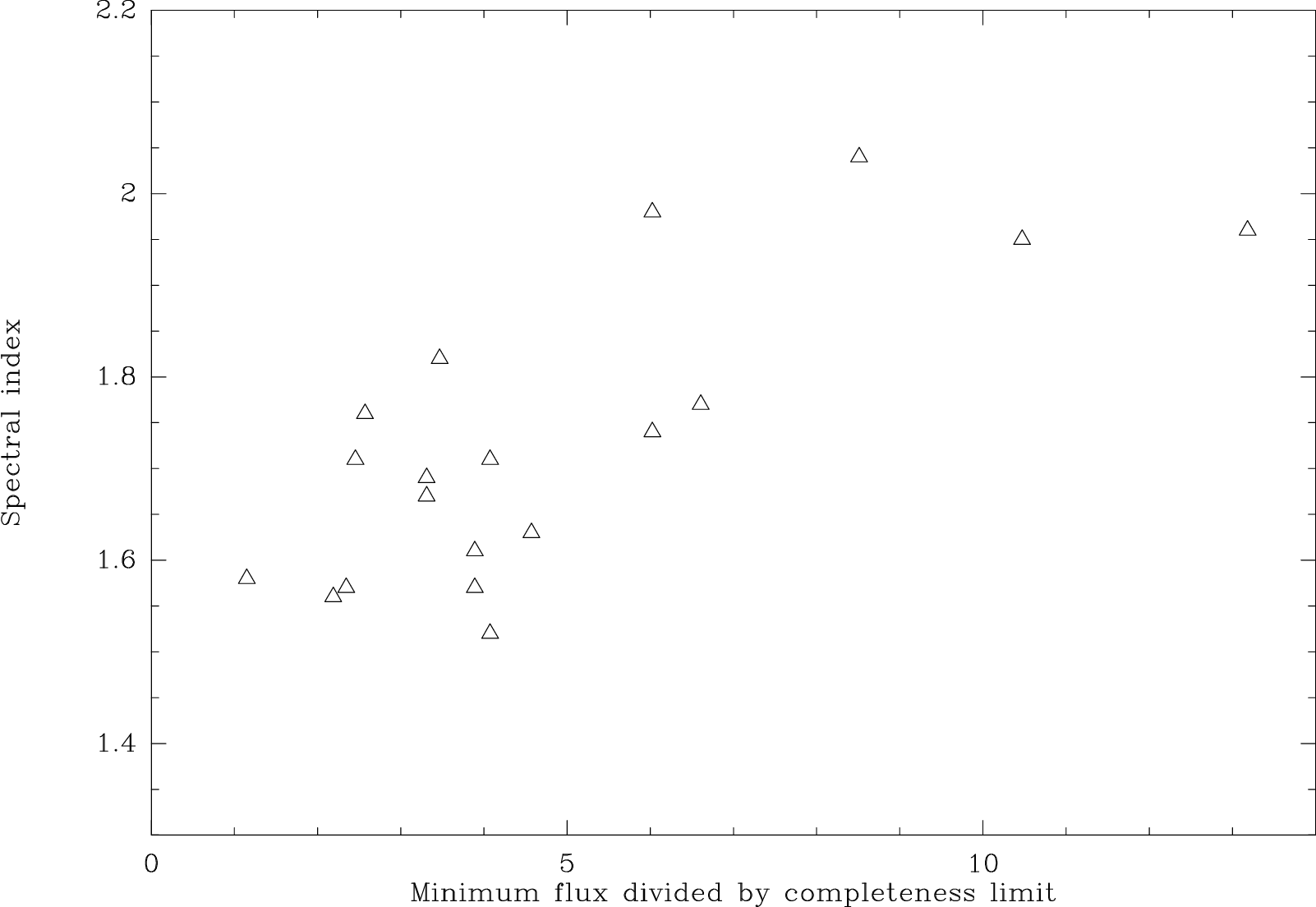}
        \caption{Link between spectral index measured with powerlaw.py and $L_{\rm min}$ for the 19 galaxies from data given in Table 2 of \citet{Santoro2022}.
    $L_{\rm min}$ is normalized by the completeness limit $L_{\rm compl}$ to be comparable from one galaxy to another.  }
        \label{Santoro} 
\end{figure}

\begin{figure}
        \centering
        \includegraphics[width=\hsize{}]{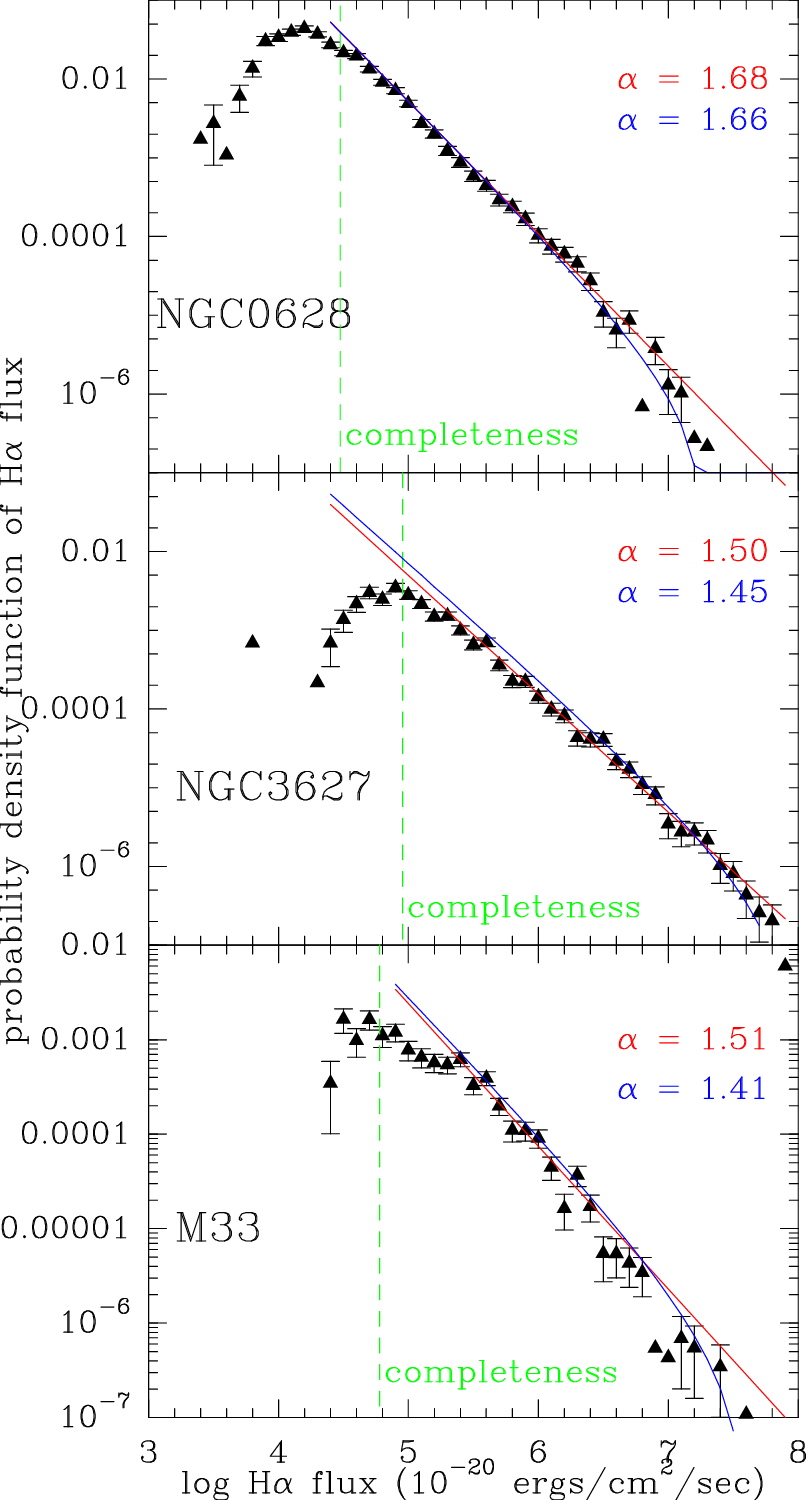}
        \caption{Probability density functions of H$\alpha$ fluxes from NGC0628, NGC3627m and M\,33, the three galaxies for which both HII region and GMC catalogs are available.  
        Results of powerlaw.py (red line) and \citet{Maschberger09} fit (blue line) along with observational data for NGC0628, NGC3627, and M33 from \citet{Santoro2022} and \citet{Lin2017} are shown.  NGC0628 and NGC3627 are the two PHANGS galaxies for which both the HII region and GMC samples are available. H$\alpha$ data for M33 are from \citet{Lin2017} The adopted completeness limit for the fits (equivalent of $L_{\rm min}$ in \citet{Santoro2022}) are shown as a green dashed line and given in Table 1.  
        The errorbars are proportional to $1/sqrt(N)$ where $N$ is the number of points in the bin. When no errorbar is plotted, it means  $N=1$.
        }
        \label{ha_pdf} 
\end{figure}

\begin{figure}
        \centering
        \includegraphics[width=\hsize{}]{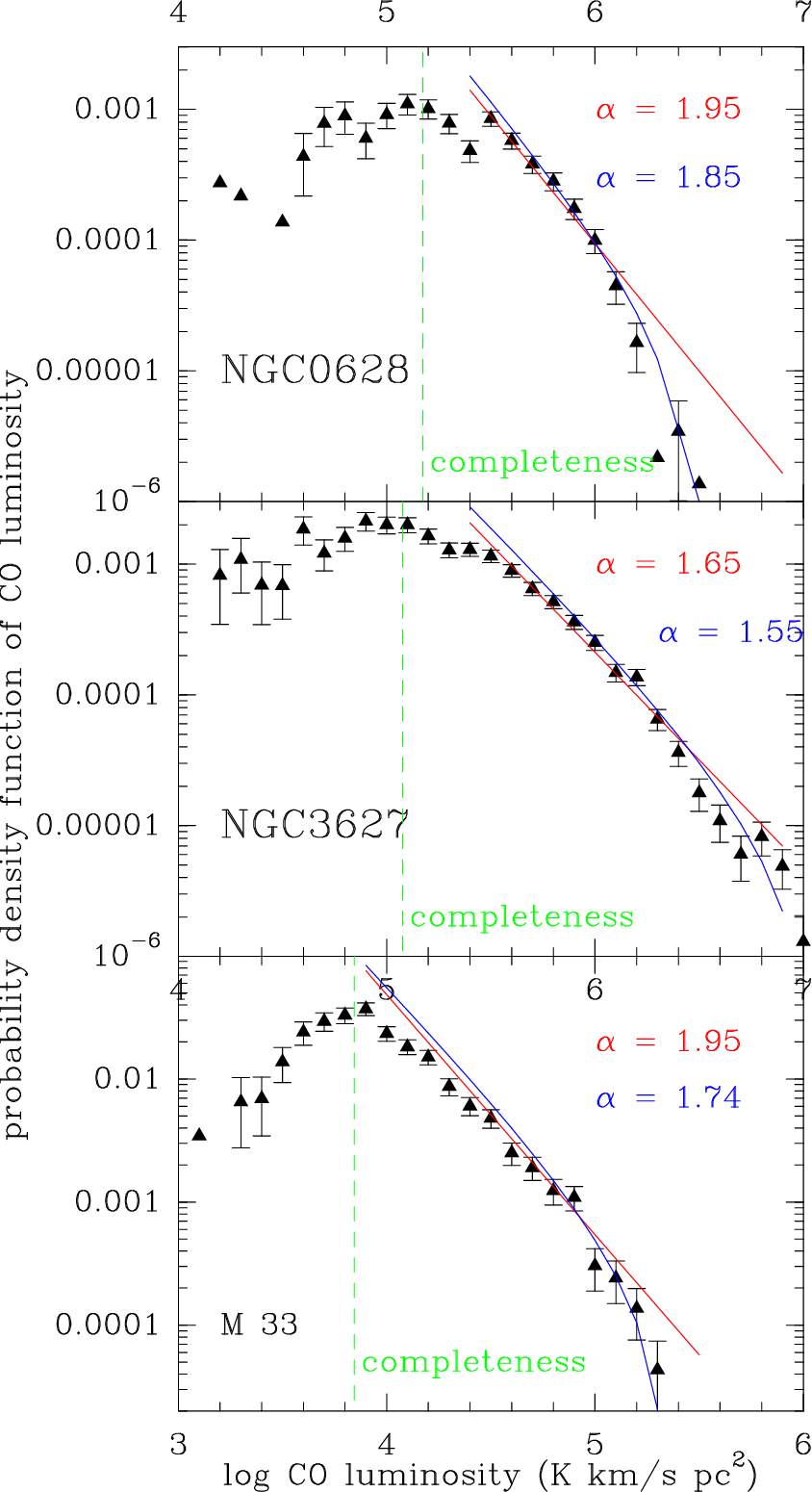}
        \caption{ Probability density functions of CO luminosities from NGC0628, NGC3627m and M\,33, the three galaxies for which both HII region and GMC catalogs are available.  Results of powerlaw.py (red line) and \citet{Maschberger09} fit (blue line) along with observational data for NGC0628, NGC3627, and M33 from \citet{Rosolowsky2021} and \citet{Corbelli17}.  NGC0628 and NGC3627 are the two PHANGS galaxies for which both the HII region and GMC samples are available. CO data for M33 are from \citet{Druard14} and \citet{Corbelli17}.
        The adopted completeness limits for the fits are the first of the limits given in Table 5 of \citet{Rosolowsky2021} and are shown as a green dashed line and given in Table 2.
        The errorbars are proportional to $1/sqrt(N)$ where $N$ is the number of points in the bin. When no errorbar is plotted, it means  $N=1$.
         }
        \label{CO_pdf} 
\end{figure}

\section{Stellar properties and methods}

\subsection{The stellar IMF}

There are three widely studied IMFs -- the original \citet{Salpeter55} IMF with the number of stars formed per interval of mass, $n(m) \propto m^{-\alpha}$ with $\alpha=2.35$, the \citet{Kroupa2001} disjoint IMF where $\alpha = 1.3$ 
for masses between 0.08 and 0.5~\msun\ but  $\alpha = 2.35$ for larger masses, and the \citet{Chabrier2003} IMF which is lognormal (see their 
Eq. 17) up to 1~\msun\ and Salpeter for higher masses.  Wherever relevant, we assume stellar masses range from 0.08 to 100~\msun.  
In practice, we use this in logarithm so $n(m)=dn/dm$ becomes (per logarithmic mass interval) $dn/d(ln(m))  = mdn/dm$, yielding \\
$dn/d(ln(m)) \propto m^{-1.35}$ for the \citet{Salpeter55} IMF \\
$dn/d(ln(m)) \propto m^{-0.3}$ for $M<M_{char}$ \citep{Kroupa2001} \\ 
$dn/d(ln(m)) \propto \frac{0.158}{ln(10)} \, e^{((log(m)-log(0.08))^2)/(2*0.69^2)}$ for $M<1~$\msun\ \citep{Chabrier2003}.  
All these IMFs have the same (Salpeter) distribution for stars with masses higher than the limiting values quoted above.

The distributions are shown in Figure \ref{imf}.  
Integrating these functions from 0.08 to 100~\msun\ yields average stellar masses of 0.28, 0.54, and 0.58 for respectively Salpeter, Kroupa, and Chabrier IMF.

\subsection{From stellar masses to luminosities}

The zero-age light-to-mass ratios, for a complete sampling, are 191, 278, and 298 \lsun/\msun\ respectively assuming a simple scaling of $L\propto M^3$.
Two more complicated but more realistic mass-luminosity relations (MLR) were also used.  One with $L\propto M^{2.3}$ for $M<0.43$~\msun, 
$L\propto M^4$ for $0.43<M<2$~\msun, $L\propto M^{3.5}$ for $2<M<55$~\msun, and $L\propto M$ for $M>55$~\msun\ approximately 
following \citet{Eker2015}. 
With this MLR, the zero-age light-to-mass ratios become 1240, 1790, and 1940 respectively, 6.5 times higher.
The second MLR uses \citet{Eker2018} Table 6 followed by \citet{Sternberg2003} for the higher masses. The main difference is that this MLR has
slightly lower temperatures and luminosities for the highest masses, yielding zero-age light-to-mass ratios of 635, 918, and 993 \lsun/\msun\ respectively.
Since this difference between the two MLRs comes from the most short-lived stars, it decreases greatly over the first Myr. 
The reader is referred to the Appendix for the details of the (IMF and) MLR.

\subsection{Populating stellar clusters}

In order to randomly sample the IMF, the standard procedure is to calculate the cumulative distribution functions of each IMF, normalize to unity, and draw random numbers uniformly between 0 and 1 and then take the star mass associated with that value of the cumulative distribution function.
Stellar clusters are built up to a given mass.  In practice the mass is slightly greater (generally less than 1\%) than the nominal cluster mass because we draw stars as long as the mass is below the nominal cluster mass.  This is reasonable as the cluster masses are far below the cloud masses so there is still a large reservoir of material left to form stars.
Random sampling results in huge luminosity variations, particularly before aging, as cluster luminosities can be dominated by a single star.  
Drawing a large number of stellar clusters naturally yields average values of the luminosities close to those calculated by directly integrating the IMF and applying the MLR.  
Clusters are populated once a cluster mass has been chosen, typically 3\% of the cloud mass, and the process is independent of the cluster or cloud mass. For a representative GMC mass distribution in the galaxies sampled in this paper we assume that GMC masses follow a power law with  $\alpha=2$, unless otherwise stated. This is the average value for the GMC distribution in PHANGS galaxies but we also show results for $\alpha = 1.7$ and $\alpha = 2.3$ to test possible effects of changing the slope of the mass distribution.

\subsection{"Randomness" of star formation}

When populating a stellar cluster by randomly sampled stellar masses following a given IMF, cluster-to-cluster luminosity variations (bolometric and H$\alpha$)
can be quite high, dependent on the mass of the most massive star. 
This is illustrated in Fig~\ref{randomness} for 1000 simulations of clusters of $\sim300$~\msun.  
If $\sim$3\% of a cloud is turned into stars, a 300~\msun\ stellar cluster 
corresponds to an initial cloud mass of $\sim$10000~\msun, approximately 
the lower limit to what is considered a GMC.
Zero-age luminosities of a 300\msun\ cluster vary by more than a factor 1000, depending on the presence of a massive star or not, 
and the dispersion in luminosities is comparable to the 
average luminosity, such that for a sample of 100 clusters major variations can be present due to the stochastic sampling of the IMF. 

For truly massive clusters, this effect becomes small although not negligible even for clusters of 10000~\msun\ of stars, as many massive stars are 
present.  This is taken into account in our simulations.

\subsection {Algorithms}

After reproducing the results presented in Figure 2 of \citet{Santoro2022}, we tested the influence of the $L_{\rm min}$ parameter and indeed found that for a given galaxy, increasing $L_{\rm min}$ above $L_{\rm compl}$ yielded a systematically steeper slope.  This is not only true for the powerlaw.py but also for the algorithms used by \citet{Rosolowsky2021} (see their Table 5).
Figure \ref{Santoro} shows how the spectral index of the galaxies depends on  
$L_{\rm min}$ normalized to the completeness limit $L_{\rm compl}$, both given in their Table 2.
When data have a physical high-end truncation coupled with increasing incompleteness at low fluxes, this behavior is expected and represents one of the difficulties.  

The powerlaw.py \citep{Alstott2014} algorithm searches for an "optimal" $L_{\rm min}$ without knowledge of $L_{\rm compl}$ but it can also be used with a fixed $L_{\rm min}$.  For each galaxy in the MUSE and ALMA samples, we ran multiple powerlaw.py fits testing many values of  $L_{\rm min}$ bracketing $L_{\rm min}$ and $L_{\rm compl}$.  While the change in slope was generally large and regular, the relative slopes between e.g. the inner and outer parts changed little.  The fact that powerlaw.py does not use knowledge of $L_{\rm compl}$, which depends on the sensitivity of the observations, is a drawback for experimental data.  \citet{Alstott2014} illustrate the powerlaw.py algorithm with word occurrence and blackout statistics which do not have a sensitivity (completeness) limit although powerlaw.py allows a minimum value to be injected.  Our approach has been to use an $L_{\rm min}$ based on $L_{\rm compl}$, equal to twice $L_{\rm compl}$ for the Santoro data and equal to $L_{\rm compl}$ for the other data.

We tried fitting using a least-squares algorithm to both the luminosities and their logarithms but this yielded visually inappropriate results.  The algorithm used by \citet{Rosolowsky2021} yielded extremely large variations in $\alpha$ for only modest changes in $L_{\rm min}$ (completeness in their Table 5) so this was not investigated further.

\citet{Gratier12} used an algorithm based on the work by \citet{Maschberger09} and this was tested on the current CO and H$\alpha$ data with reasonable results, generally similar to powerlaw.py for similar $L_{\rm min}$.  
Figures \ref{ha_pdf} and \ref{CO_pdf} present results using both powerlaw.py and the \citet{Maschberger09} technique.
The Maschberger-Kroupa power-law fit uses a truncated power law. The powerlaw.py algorithm has no truncation. 

On simulated data taken from a power law distribution ($\alpha = 1.4, 1.5...2.0$) with random added noise, both powerlaw.py and the  
\citet{Maschberger09} algorithms performed well although for a given noise level, the  \citet{Maschberger09} did 
measurably better in recovering the initial distribution.  

\section{HII region and GMC luminosities across galaxy disks}

We calculate the Probability Density Function (hereafter PDF) as the number of objects per bin divided by the bin width.  For example, at log(CO luminosity) $=6$ and for a bin width of 0.1 dex as in Fig. \ref{CO_pdf}, the width is $2.3\, \, 10^5$ so if 20 objects are present, the PDF for that luminosity bin is $20/2.3e5 = 8.7 \, \, 10^{-5}$.  Given that the PDFs are presented per galaxy, such that the distance to each region is the same, the shape of a flux PDF or a luminosity PDF is the same.  The slope $\alpha$ of the fit to the PDF is the spectral index.

Figure \ref{ha_pdf}  shows the PDF of the HII region H$\alpha$ fluxes of NGC0628, NGC3627, and M33.  
The powerlaw.py fit is shown as a red line and the \citet{Maschberger09} fit in blue, with the assumed completeness level (used for both fits) 
indicated with a green dashed line.
These galaxies have been chosen because they are the only PHANGS galaxies for which both 
HII region and GMC catalogs are available.
Figure \ref{CO_pdf} shows the results for CO luminosities of the same galaxies.  It can be seen that the CO results are based on a significantly smaller 
dynamic range (ratio between completeness level and maximum) than the H$\alpha$ flux distribution.

In order to fit the GMC distributions, we have chosen the lower completeness level given by \citet{Rosolowsky2021} in their Table 5 and reproduced in Table 2. The GMC slopes are clearly steeper than for the HII regions.
The GMC completeness limit was originally 
given by \citet{Rosolowsky2021} as a mass so we have 
converted this to a CO luminosity using their Eq. 5 for a solar metallicity.
We use the CO luminosities because they are the observed quantities (whereas the mass requires a conversion factor) and directly comparable to the H$\alpha$ luminosities.  
The fits were also performed using the masses in the online table with no significant difference.  See Section 5.1 for a discussion of the effect of a change in $\ratioo$.

The fitting results are given in Table 1 
for the HII regions and Table 2 
for the GMC sample.  
The fits using the powerlaw.py algorithm \citep{Alstott2014} and the \citet{Maschberger09} procedure are shown as $\alpha_{\rm py}$ and $\alpha_{\rm MK}$ respectively.
The uncertainties are estimated via the bootstrapping method with 100 random draws for each fit.
For the HII regions, we find $1.47 \le \alpha_{\rm py} \le 1.87$ (average 1.64 and dispersion 0.10) and $1.42 \le \alpha_{\rm MK} \le 1.86$  (average 1.61 and dispersion 0.11). 
Linking the minimum flux used for the fits to the completeness level given in Table 2 of \citet{Santoro2022} results in a decrease of the dispersion from 0.15 to 0.10.
For the GMCs, we find $1.65 \le \alpha_{\rm py} \le 2.98$ (average 2.09 and 
dispersion 0.42) and $1.55 \le \alpha_{\rm MK} \le 3.00$ (average 2.14 and dispersion 0.39).  The dispersion in $\alpha$ found by \citet{Rosolowsky2021} using different methods is considerably higher (1.14 and 0.44 from  cols 3 and 6 of their Table 5) for the 8 galaxies with more than 100 GMCs.
Our two fitting methods are in good agreement for both the HII regions and the GMCs, showing that the average spectral index $\alpha_{\rm CO} > \alpha_{\rm H\alpha}$. For the three galaxies in common, NGC 628 and NGC 3627 in the PHANGS sample and M33, the slope of the GMC PDF is also steeper than that of the HII region PDF so this is likely to be a general feature. 

\begin{figure}
        \centering
        \includegraphics[width=\hsize{}]{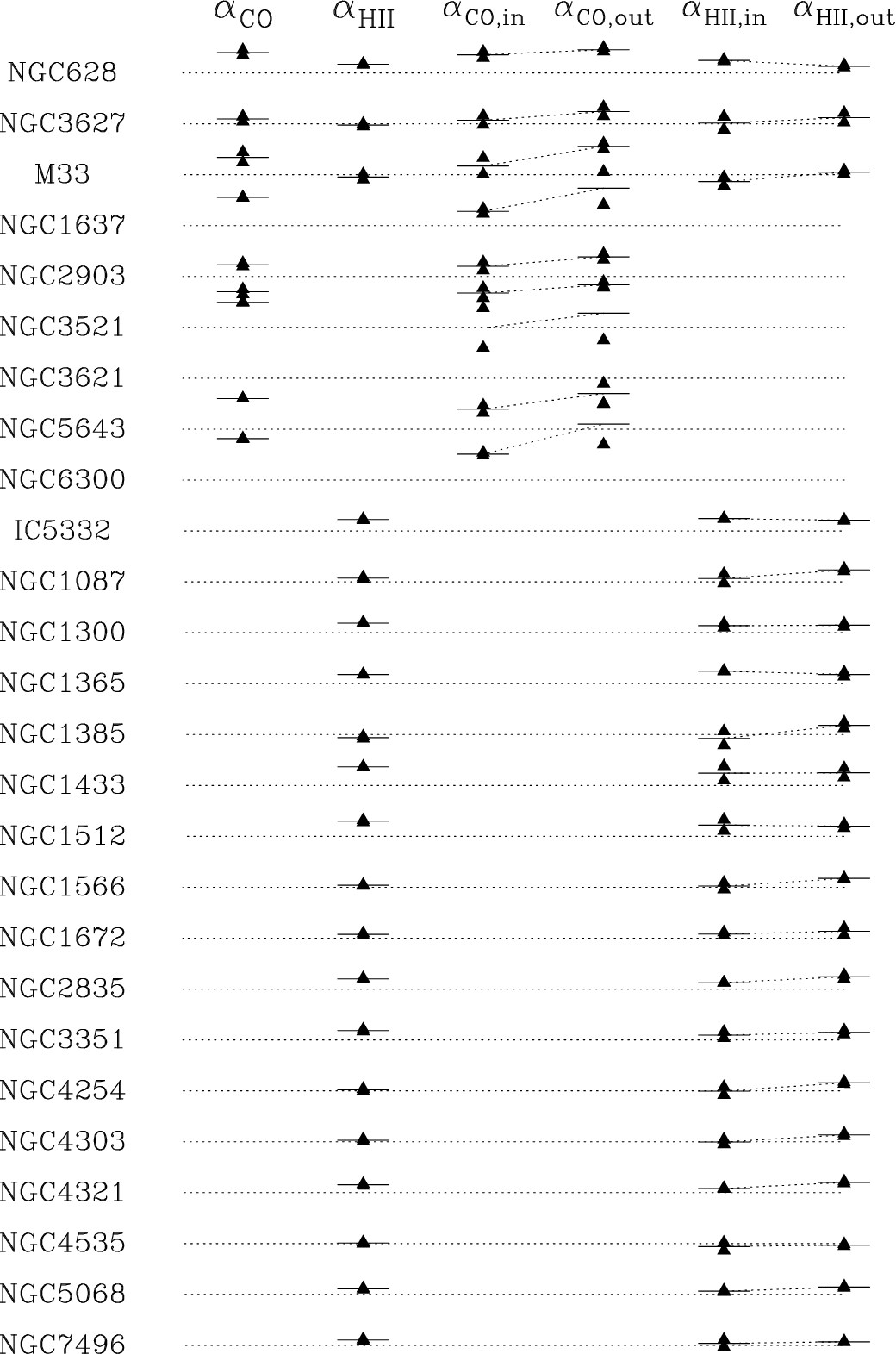}
        \caption{ Results of fits to the distribution of H$\alpha$ and GMC CO luminosities of star forming regions.  For each galaxy, the name is indicated followed by (1) the whole-galaxy GMC spectral index
        (2) the whole-galaxy H$\alpha$ spectral index (3) the inner and outer GMC spectral index, connected by a line, and (4) the same for the H$\alpha$ luminosities of HII regions.  In each case, the 
        triangles indicate the values obtained from the two independent fitting methods and the average by the short horizontal line.  
        When the line goes up, the distribution in the outer part is steeper (relatively more small objects). A horizontal dashed line at a value of 1.5 is plotted to allow 
        easy appraisal of the values. }
        \label{alpha} 
\end{figure}

\begin{figure}
        \centering
        \includegraphics[width=\hsize{}]{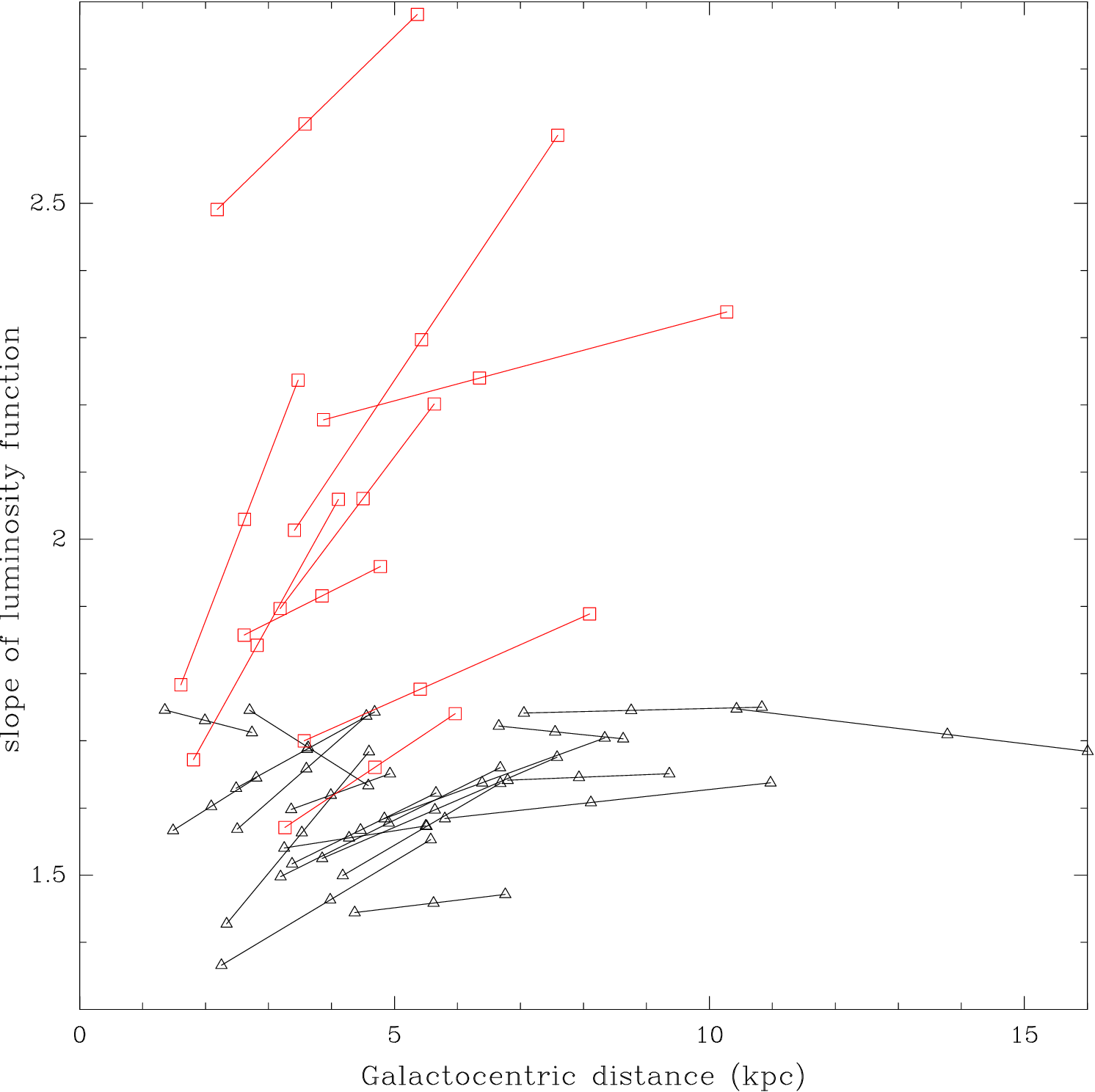}
        \caption{Radial variation of spectral indices.  This plot portrays the latter 4 columns of Fig.~\ref{alpha} in a simpler fashion. 
        The lines start with the median position of the inner regions (HII or GMC) and end at the median position of the outer regions.
        The marker near the line center indicates the median galactocentric distance of the whole sample and 
        column 5 of Tables 1 and 2
         provides the correspondance with the optical radius R$_{25}$.
        Red lines are GMC indices and black those of HII regions.
        When the line goes up, the spectral index in the outer part is higher because the distribution is steeper (relatively more small objects).  }
        \label{alpha_rad} 
\end{figure}

Figure \ref{alpha} attempts to illustrate the slopes and their radial variation.  
For each galaxy, with the galaxies having both GMC and HII region data at the top, 
the CO index $\alpha_{\rm CO}$ is shown followed by the HII region index $\alpha_{\rm H\alpha}$.  Both fits are presented (triangles) and the horizontal bar is the average. 
When only a single triangle is visible, it simply means the agreement between the methods was excellent.  The horizontal dotted line is meant as a 
guide and shows a slope of 1.5, making it clear that the $\alpha_{\rm CO}$ are steeper then $\alpha_{\rm H\alpha}$.  
The following columns show how the spectral index $\alpha$ varies between the inner and outer parts.  
All CO luminosity functions 
become steeper, in agreement with previous work.  A majority of the H$\alpha$ luminosity functions 
steepen with radius but less so and with several cases where they do not.
Figure~\ref{alpha_rad} illustrates the radial variation of $\alpha$ in a simpler fashion.

The general features are that  (1) the CO luminosity functions are steeper than the H$\alpha$  luminosity functions and (2) the distribution of CO luminosity
steepens systematically in the outer parts, unlike the distribution of H$\alpha$ luminosity.  
This is independent of the fitting routine and of any reasonable choice of parameters so we must now explore possible reasons.

\section{Why is $\alpha_{\rm CO}$ steeper than $\alpha_{H\alpha}$ ?}

Intrinsically, for a given IMF, we expect that a cloud will convert some fraction (2-3\%) of its mass into stars and hence create an HII region 
of a luminosity which should depend on the initial cloud mass.  At some point the cloud will be dispersed by supernovae (SN) and winds.  
Thus one would expect that, at least before the first SN, that the slopes of the luminosity functions should be the same.

The GMC samples discussed here have cloud masses well in excess of 10,000~\msun.  
The HII regions are also quite luminous so we will only be discussing HII regions with initial stellar masses 
well in excess of the mass of an individual massive star, assumed to be $\le100$~\msun.  
While there are examples of star formation in low-mass clouds, in GMCs stars tend to form in dense 
cores ($<$pc), themselves within dense clumps ($1-10$~pc) within the GMC.  
The clouds studied here are GMCs ranging from $10^4$ to over $10^6$\msun\ so clusters rather than individual stars are formed and the mass reservoir is large even if only 3\% of the GMC mass is converted into stars.

It appears that the power-law part of the IMF, i.e. beyond about 2~\msun (see Fig.~\ref{imf}), is due to different processes and 
independent of whether the lower end follows a Salpeter or Kroupa or Chabrier distribution. 
A high local gas density and/or low metallicity could push up the characteristic mass (manifested by the 
turnover of the Kroupa IMF) but probably not affect the high-mass end \citep[see discussions in][]{Larson1998,Kroupa2001,Hopkins2018}.  
Since the HII region luminosity is completely dominated by the high-mass end of the IMF, these changes should not affect the link between cloud mass and HII region luminosity.

In the rest of this Section we consider effects which complicate this picture such as: \\
-- The conversion from CO luminosity to H$_2$ mass is not a constant. \\
-- The luminosity of a given HII region depends strongly on the mass of the most massive star such that a cluster containing 1000~\msun\ of 
stars drawn randomly can have a radically different luminosity depending on whether a truly massive star is "drawn".  
This adds noise to the system and is why it is important to have many regions.\\
--HII regions age quickly in that their population of massive stars changes significantly over the lifetime of a molecular cloud 
($\sim 15$~Myr, approximately the lifetime of a O9 or B0 star).  \\
-- The IMF could vary.  An interesting proposal we are aware of is a link between the mass of the most massive star $M_{*,max}$ in the cluster and 
the parent molecular cloud mass as proposed by \citet{Larson1982}.  

\subsection{Variations of $\ratioo$ with galactocentric radius}

Metallicities tend to decrease with galactocentric radius, as do gas temperatures, and both tend to result in an increase of the $\ratioo$ conversion
further from galactic centers.  The basis for the $\ratioo$ factor is clearly presented in \citet{Dickman86} and recent discussions of the $\ratioo$ factor can be found in \citet{Teng2024}, \citet{Schinnerer2024}, and \citet{Leroy2025}.  We made tests by randomly choosing positions and drawing cloud CO luminosities with a predefined spectral index $\alpha$ 
and then added a radial $\ratioo$ gradient to obtain cloud masses.  The data were then treated as the real data but the $\alpha$ determined 
is not affected by the$\ratioo$ because $\ratioo$ depends on the radial distance but not on the cloud luminosity or mass.
So any $\ratioo$ variation does not affect $\alpha$ unless $\alpha$ depends on the cloud mass or luminosity. A radial variation affects 
clouds of all masses in the same way.

\subsection{Aging of HII regions}

A bit like the $\ratioo$ factor, if all masses are affected in the same way, which would be the case if the IMF is not affected by the parent cloud mass, 
then this should not affect $\alpha_{H\alpha}$.  However, if only massive clouds form truly massive stars, then these will disappear quickly such that the
greatest declines in luminosity (as a fraction of the original $t=0$ luminosity) will be for the HII regions created in massive GMCs, as shown in Fig.~\ref{starlife} for the three commonly used IMFs.  

The rapid death of the most massive stars implies that a large sample is necessary and that comparisons must be made over the full life cycle
of an HII region.  Fig.~\ref{clust_sim} shows how aging affects cluster bolometric and H$\alpha$ luminosities. 
Each cluster is assigned a uniformly random age between 0 and 15~Myr, which is when the cluster emits less than 1\% of the original number of ionizing photons.
The $t=0$ cluster luminosity distribution is created based on randomly generated stellar masses. Cluster masses follow the mass distribution indicated: solid, dashed, or dotted lines for $\alpha_{cl}=2.3$, 2.0, or 1.7 respectively.
Fig.~\ref{clust_sim} shows the $t=0$ luminosity function in the lower panel) and with   aging in the upper panel.
The Chabrier IMF has been used.  Fig.~\ref{imf} and the Appendix show how the IMFs affect cluster properties.  Briefly, the Kroupa and Chabrier IMFs are extremely similar and the Salpeter IMF has more low-mass stars.
For each age, the cluster luminosity (or number of ionizing photons) is only integrated up to the most massive star still present 
(see bottom panel of Fig~\ref{starlife}).  The resulting luminosity distribution is then recalculated.  
The same is done for the number of ionizing photons.  
Fig.~\ref{clust_sim} provides the H$\alpha$ luminosity generated by the ionizing photons, under the simple assumption that each photo-ionized H atom produces an H$\alpha$ photon.  The H$\alpha$ fluxes generated cover the range observed by \citet{Santoro2022} in their Fig. 2. 

Randomness has some curious effects.  The bottom panel of Fig.~\ref{clust_sim}  shows the results of simulating $3 \times 30000$ (see caption for details) clusters between 100 and 10000~\msun. The slopes of the initial mass function were 
$\alpha_{cl}= 1.7$, 2.0, and 2.3 and then the clusters were binned by luminosity. 
The randomness in drawing 
a massive star implies that the luminosity distribution is quite unlike the mass distribution. 
 And there is only a very short high luminosity power-law tail.
The difference between the slopes of the mass and of the luminosity distributions is clearly seen for the high luminosities. 
This changes significantly when allowing for aging.  
The simulations are done for three different cluster mass functions and each value of $\alpha_{cl}$ has 90000 clusters 
simulated. 
The results with aging are shown in the top panel of Fig.~\ref{clust_sim} and while the luminosity distribution is not a power law 
(this can be seen by the convex shape, particularly around bolometric luminosities $L\sim 10^5 - 10^6$~L$_\odot$), there is a monotonic decrease which is close to a power law distribution.  The slopes of the bolometric and H$\alpha$ luminosity functions are somewhat shallower than the cluster mass function (note the difference between the luminosity and the mass x-scales).

\subsection{Testing an IMF where $M_{*,max}$ depends on the cloud mass}

We then directly tested the  \citet{Larson1982} link $M_{*,max} = 0.33 M_{\rm GMC}^{0.43}$ by generating a mock set of 
cluster masses assuming that 3\% of the cloud mass was converted into stars ($M_{\rm clust} = 0.03 \times M_{\rm GMC}$).  
The cluster masses range from 100 to 10000~\msun.  The mass functions tested has again  $\alpha_{cl}= 1.7$, 2.0, and 2.3.
Because the smaller clouds form lower mass massive stars, their H$\alpha$ luminosity is decreased with respect to
large clouds, resulting in a larger range of H$\alpha$ luminosities and hence a shallower slope, very much like what is observed.

\begin{figure}
        \centering
        \includegraphics[width=\hsize{}]{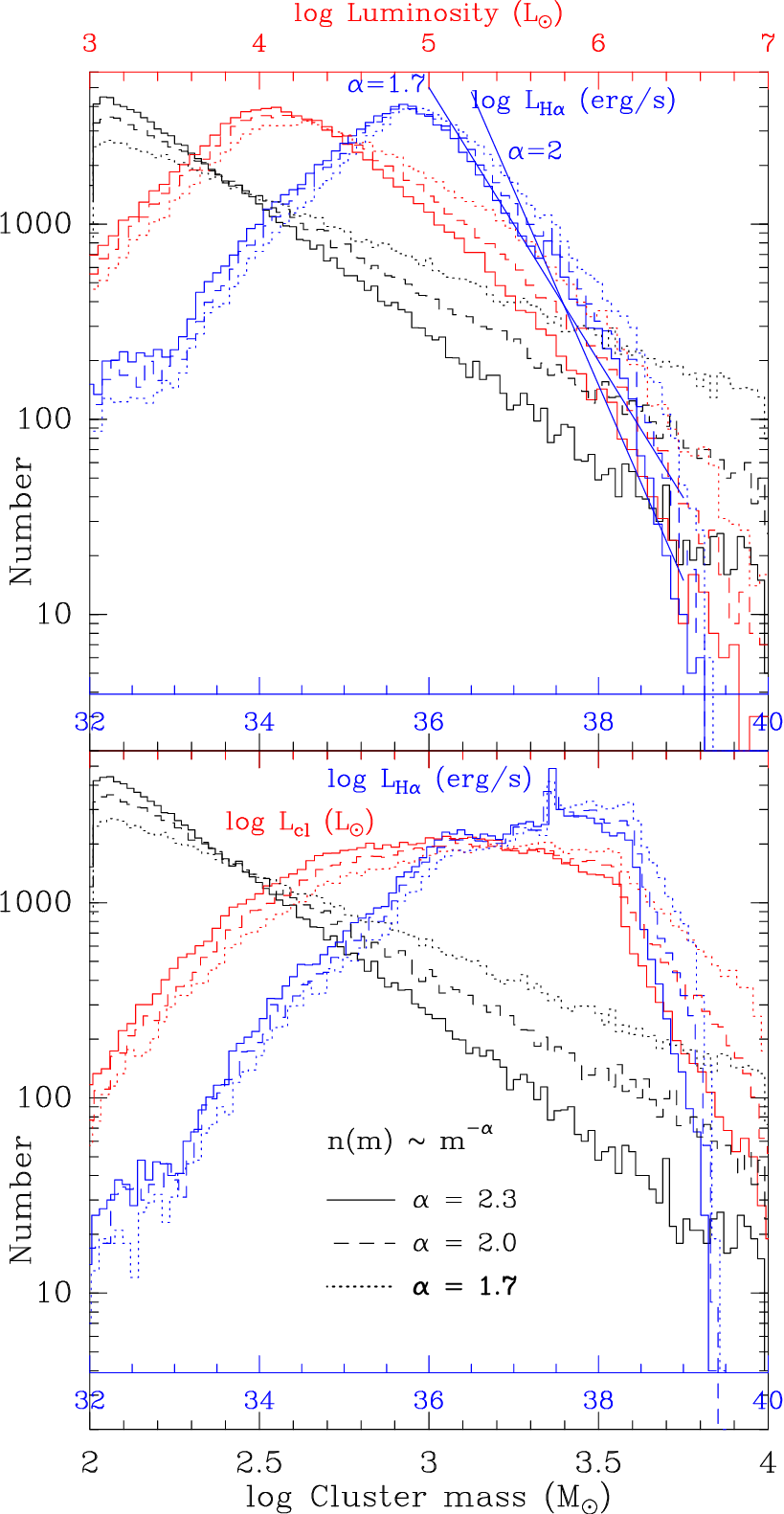}
        \caption{Histogram of simulated cluster properties assuming a "stochastic" IMF. 
       Black lines represent the cluster masses, red their luminosities, and blue the H$\alpha$ luminosity.  The zero-age IMF is shown in the lower panel) and with aging as described in the text in the upper panel.
        The solid-dashed-dotted lines, of whatever color, show the results for the different cluster mass functions ($\alpha_{\rm cl}=2.3$, 2.0, 1.7).
        Similarly, the red x-axis provides the cluster luminosities and the blue x-axis the H$\alpha$ luminosity, both in logarithm. In the upper panel, two blue lines are shown with slopes of 1.7 and 2.0 to help guide the eye.
        We drew 3 sets of 30000 clusters for each of the mass functions and, after checking that they yielded the 
        same results, they were combined to further reduce any noise so the distributions are based on 90000 clusters for each mass function.
        The Chabrier IMF was used with the Eker-Sternberg MLR.}
        \label{clust_sim} 
\end{figure}

\begin{figure}
        \centering
         \includegraphics[width=\hsize{}]{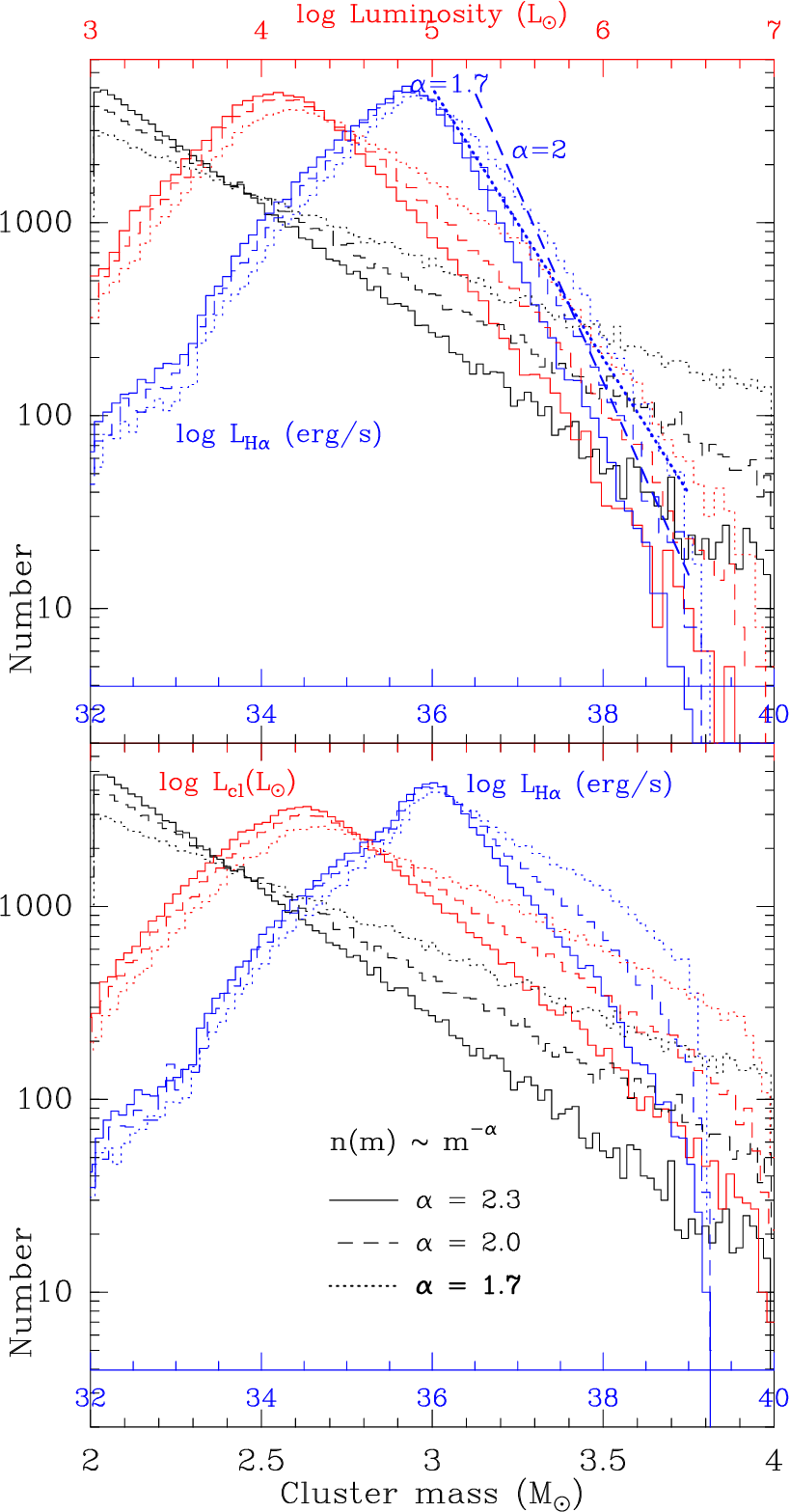}
        \caption{ Like Figure~\ref{clust_sim} (zero-age in lower panel, random aging up to 15Myr in upper panel) 
        but where the maximum stellar mass is 
        as proposed by \citet{Larson1982}, see Sect. 5.3. One of the main differences is the lack of the broad peak near $10^{38}$erg/s.  The peak is due to the presence of a massive star in a small cluster, no longer possible with the \citet{Larson1982} mass limit.
While the cluster mass distribution is indeed a power law, the cluster luminosity and $N_{ion}$ function has 
       an $\alpha$ which increases slowly with the minimal luminosity used in the fit, whether the Maschberger or powerlaw.py algorithm is used. 
         This is seen as the curvature not present in the cluster mass distribution and is the result of the
         limiting stellar mass. }
        \label{gmc2hii2} 
\end{figure}

\begin{figure}
        \centering
        \includegraphics[width=\hsize{}]{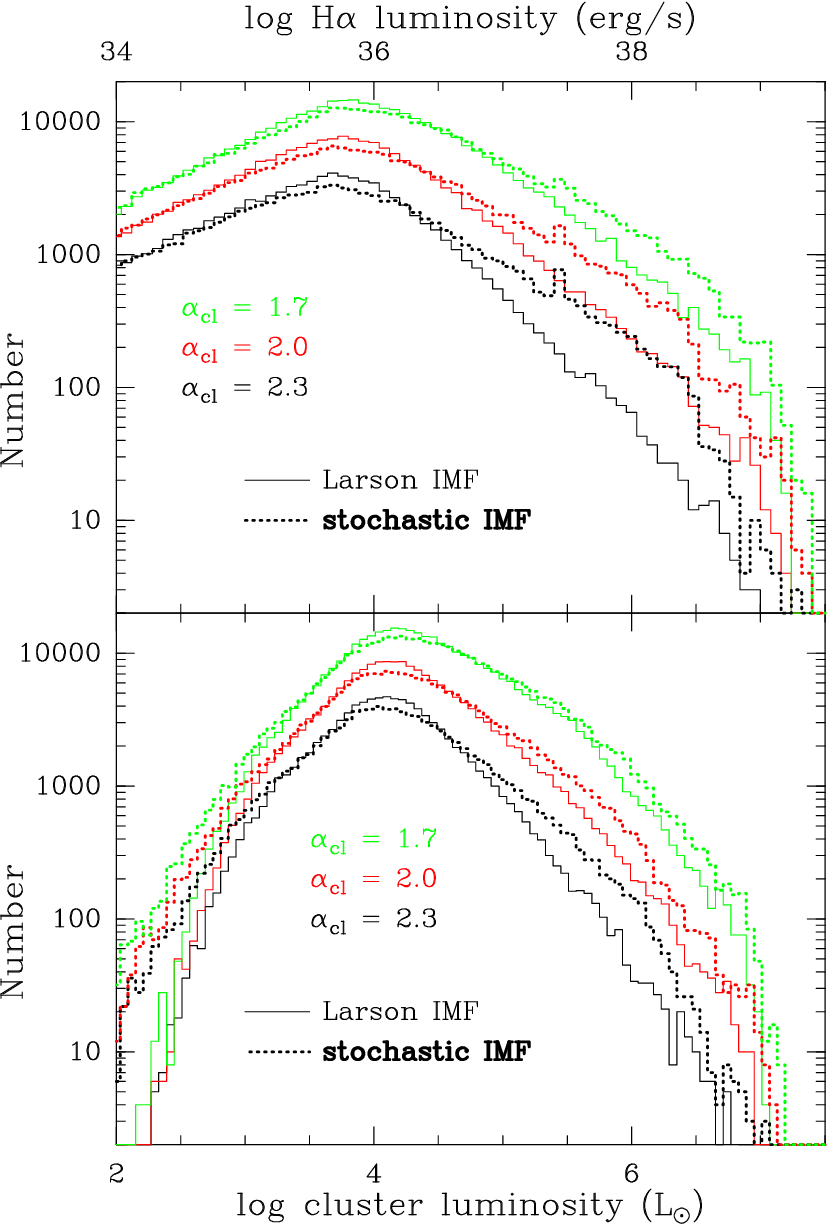}
        \caption{Comparison of the distribution of cluster properties for a stochastic IMF (0.08 -- 100\msun) with that derived for Larson's model \citep{Larson1982} where
        the maximum stellar mass varies with the parent cloud mass.  
        The cluster mass distribution is a pure power law as in the previous figures with $\alpha_{cl} = 1.7$, 2.0, and 2.3 and $M_{\rm clust} = 0.03 M_{\rm cl}$.
        Cluster  ages are attributed randomly as described in the text.  The bolometric luminosities (bottom) and H$\alpha$ luminosity (top) are 
        then calculated for the stellar population at the age of each HII region. This figure can be directly compared with Fig.~\ref{gmc2hii2}.  }
        \label{gmc2hii_age} 
\end{figure}

Fig.~\ref{gmc2hii2} shows the zero-age HII region bolometric and H$\alpha$ luminosity distributions, as in Fig.~\ref{clust_sim} but with the \citet{Larson1982} $M_{*,max}$.  The initial slopes of the luminosity and H$\alpha$ distributions are significantly shallower than the parent 
GMC distribution and have a slightly convex shape.  However, after aging, the slopes are no longer significantly 
shallower than the cluster mass distribution.  The comparison between the stochastic (full 0.08 -- 100\msun\ range) and \citet{Larson1982} limited IMFs is shown in Fig.~\ref{gmc2hii_age} 
where it is apparent that the \citet{Larson1982} stellar mass limit yields luminosity functions that are steeper than the stochastic IMF covering the full mass range.

\begin{figure}
        \centering
        \includegraphics[width=\hsize{}]{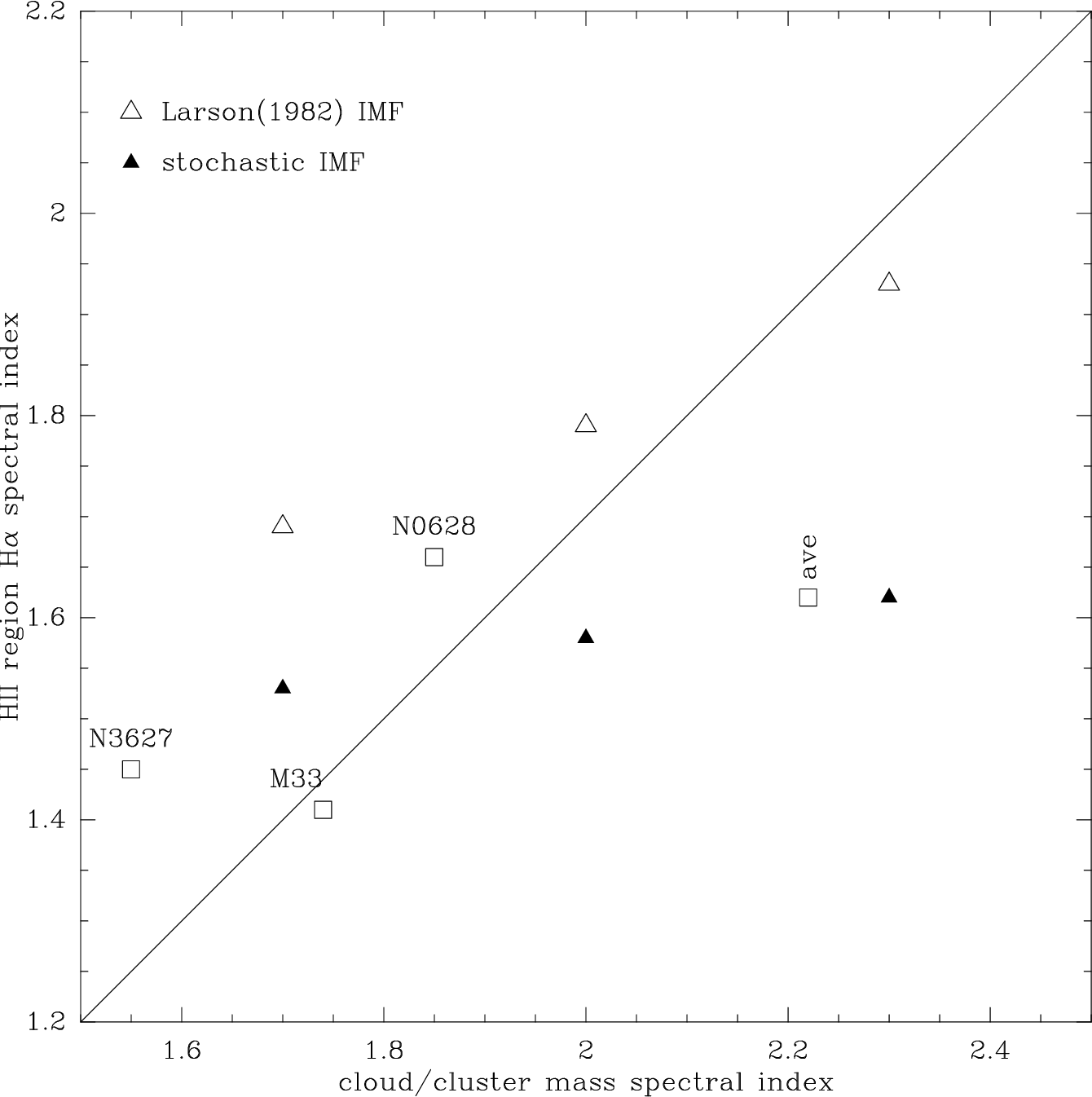}
        \caption{Comparison of observed GMC spectral index and HII region H$\alpha$ luminosity spectral index with the simulations using the 
        stochastic and \citet{Larson1982} limited IMFs.  The point marked "ave" represents the average of the galaxies observed in H$\alpha$ but not in CO plotted with the average of the galaxies observed in CO but not H$\alpha$.
       }
        \label{gmc2hii2fig} 
\end{figure}

The difference in slope is shown in Fig.~\ref{gmc2hii2fig} where the values for the 3 galaxies with both measured slopes as well 
as the average  $\alpha_{\rm CO}$ and $\alpha_{\rm H\alpha}$ for the remaining galaxies (excluding NGC0628, NGC3627, and M33) are plotted.  
The open squares show $\alpha_{\rm cl}$ and $\alpha_{\rm H\alpha}$ from the simulations given in Table 3 using the MK fit.
The powerlaw.py fit (also provided in Table 3), for the same minimum value, yields slopes about 2\% steeper.  These are simulations that take into account aging (upper panels of Figures \ref{clust_sim} and \ref{gmc2hii2}).   

When $\alpha_{\rm cl}=2$ (our canonical case corresponding 
to the average of the PHANGS$+$M33 sample), the difference in slopes $\alpha_{\rm cl}$--$\alpha_{\rm H\alpha}$ is about 0.2 for the Larson mass limit and 0.4 for the stochastic star formation (see Table 3), 
which is closer to the observed values in Figure~\ref{gmc2hii2fig} .

\begin{figure}
        \centering
        \includegraphics[width=\hsize{}]{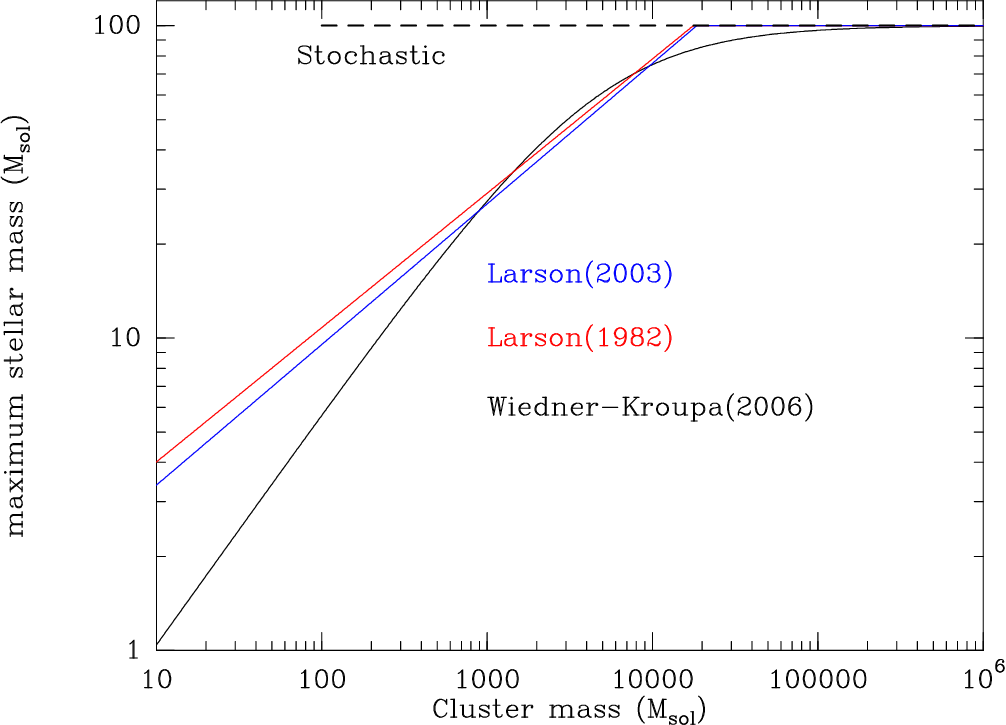}
        \caption{  Comparison of the \citet{Larson1982}, \citet{Larson2003}, and Kroupa-Weidner \citep{Weidner2004,Weidner2006} maximum stellar masses as in \citet{Weidner2006} Fig. 1 but recalculated for the range in stellar masses used here (0.08--100~\msun) and a Chabrier IMF.  The dashed line at 100\msun gives the "stochastic" mass limit.
       }
        \label{Larson-Weidner} 
\end{figure}

Two options have been examined here -- the stochastic and the \citet{Larson1982} maximum stellar mass.  However, \citet{Weidner2006} also proposed a limiting mass for low 
mass clusters and \citet{Larson2003} presented a somewhat modified version of his 1982 proposal. In Fig. \ref{Larson-Weidner} the maximum stellar masses predicted by the various theories are compared as in Fig. 1 of \citet{Weidner2006} except that the values have been recalculated for an IMF going from 0.08\msun\ to 100\msun\ using a Chabrier IMF (very similar to the Kroupa IMF -- see Fig. \ref{imf}) in order to make them comparable to our study.  The other stellar mass limits proposed give more discrepant results from stochastic sampling than the \citet{Larson1982} maximum mass, and hence cannot improve the agreement of the simulated distributions with the observed spectral indices in Fig. \ref{gmc2hii2fig}. 

Intuitively, a stellar mass limit below the top of the IMF results in more low luminosity HII regions, resulting in a steeper slope when fit with a power law.  The lower the limit, the stronger the effect. 

\begin{table*}
\caption[]{H$\alpha$ results for PHANGS MUSE galaxies}
\begin{center}
\begin{tabular}{lllllllllll}
Galaxy & $\alpha_{MK}$&$\alpha_{py}$ & n$_{\rm HII}$ & $F_{min,H\alpha}$ & \multicolumn{2}{c}{R$_{\rm med}$} & $\alpha_{MK,in}$,$\alpha_{py,in}$ & n$_{\rm HII,in}$ & $\alpha_{MK,out}$,$\alpha_{py,out}$ & n$_{\rm HII,out}$  \\
\hline
IC5332 & $1.72\pm 0.033$&$1.74\pm 0.020$ & 476 & 11000 & 1.99 &0.25& 1.74,1.75 & 258 & 1.70,1.73 & 218 \\
NGC0628 & $1.66\pm 0.016$&$1.68\pm 0.017$ & 1421 & 30002 & 3.63  &0.26& 1.73,1.76 & 650 & 1.61,1.66 & 771 \\
NGC1087 & $1.56\pm 0.021$&$1.60\pm 0.013$ & 588 & 50476 & 3.60 &0.53& 1.48,1.66 & 371 & 1.76,1.71 & 217 \\
NGC1300 & $1.68\pm 0.025$&$1.71\pm 0.020$ & 745 & 20000 & 7.93 &0.48 & 1.68,1.60 & 398 & 1.68,1.62 & 347 \\
NGC1365 & $1.68\pm 0.032$&$1.69\pm 0.027$ & 419 & 100000 & 13.78 &0.40 & 1.74,1.76 & 185 & 1.64,1.73 & 234 \\
NGC1385 & $1.42\pm 0.020$&$1.47\pm 0.013$ & 710 & 26982 & 3.53 &0.41 & 1.29,1.57 & 438 & 1.74,1.63 & 272 \\
NGC1433 & $1.86\pm 0.036$&$1.87\pm 0.035$ & 588 & 30000 & 8.76 &0.52 & 1.88,1.60 & 292 & 1.84,1.66 & 296 \\
NGC1512 & $1.78\pm 0.042$&$1.82\pm 0.040$ & 324 & 30000 & 7.55 &0.33 & 1.83,1.61 & 163 & 1.74,1.67 & 161 \\
NGC1566 & $1.53\pm 0.014$&$1.55\pm 0.013$ & 1078 & 31455 & 5.64  &0.30 & 1.46,1.59 & 665 & 1.68,1.67 & 413 \\
NGC1672 & $1.56\pm 0.020$&$1.60\pm 0.021$ & 694 & 47592 & 8.12 &0.47 & 1.55,1.62 & 371 & 1.57,1.71 & 323 \\
NGC2835 & $1.69\pm 0.030$&$1.72\pm 0.033$ & 488 & 52360 & 3.61&0.32  & 1.63,1.63 & 267 & 1.78,1.71 & 221 \\
NGC3351 & $1.67\pm 0.024$&$1.71\pm 0.020$ & 670 & 19347 & 3.99 &0.38 & 1.66,1.54 & 355 & 1.69,1.61 & 315 \\
NGC3627 & $1.45\pm 0.017$&$1.50\pm 0.017$ & 833 & 90250 & 4.46 &0.26 & 1.39,1.64 & 433 & 1.53,1.72 & 400 \\
NGC4254 & $1.50\pm 0.013$&$1.54\pm 0.007$ & 1847 & 57359 & 4.91&0.51 & 1.42,1.58 & 1106 & 1.64,1.68 & 741 \\
NGC4303 & $1.51\pm 0.014$&$1.55\pm 0.009$ & 1613 & 48167 & 5.51&0.32  & 1.45,1.55 & 961 & 1.61,1.66 & 652 \\
NGC4321 & $1.64\pm 0.020$&$1.67\pm 0.014$ & 1005 & 61501 & 6.39&0.47  & 1.59,1.58 & 560 & 1.72,1.69 & 445 \\
NGC4535 & $1.50\pm 0.013$&$1.53\pm 0.010$ & 1260 & 9067 & 5.62&0.30  & 1.52,1.37 & 629 & 1.49,1.45 & 631 \\
NGC5068 & $1.60\pm 0.021$&$1.62\pm 0.017$ & 930 & 44785 & 2.09&0.37 & 1.58,1.55 & 471 & 1.63,1.66 & 459 \\
NGC7496 & $1.59\pm 0.032$&$1.62\pm 0.017$ & 377 & 20822 & 4.28&0.47  & 1.61,1.47 & 213 & 1.58,1.57 & 164 \\
M 33     & $1.41\pm 0.026$&$1.51\pm 0.018$ & 368 & 60000  & 3.98$^a$ &0.53 & 1.28,1.45 & 190 & 1.52,1.59 & 178 \\
\hline
\end{tabular}
\tablefoot{Sample of galaxies observed by \citet{Santoro2022} followed by results from fitting the H$\alpha$ flux distribution with the \citet{Maschberger09} method and powerlaw.py algorithm \citep{Alstott2014}. 2nd column is overall slope, where the second number is determined with powerlaw.py. 3rd column is the number of HII regions with a flux greater than the value in col. 4 expressed in units of $10^{-20}$ergs/cm$^{-2}$/sec. Col 4 is designed to be the true completeness limit as explained in the text. Col. 5 gives the median galactocentric radius of the HII region sample for that galaxy in kpc and as a fraction of R$_{25}$.  The following two columns provide the inner slope with the two methods and the number of HII regions. The next two columns are the same but for the outer parts.  $^a$ Corrected for a distance of 840 kpc. Due to the lower number of HII regions, the uncertainties on $\alpha$ are higher when the inner and outer parts are evaluated separately.}
\end{center}
 \label{ahii} 
\end{table*}

\begin{table*}
\caption[]{CO results for PHANGS galaxies}
\begin{center}
\begin{tabular}{lllllllllll}
Galaxy & $\alpha_{MK}$&$\alpha_{py}$ & n$_{\rm GMC}$ & $L_{min,CO}$ &  \multicolumn{2}{c}{R$_{\rm med}$} & $\alpha_{MK,in}$,$\alpha_{py,in}$ & n$_{\rm GMC,in}$ & $\alpha_{MK,out}$,$\alpha_{py,out}$ & n$_{\rm GMC,out}$  \\
\hline
NGC0628 & $1.85\pm 0.040$&$1.95\pm 0.030$ & 368 & 149254 & 3.85&0.27 & 1.80,1.92 & 205 & 1.92,2.00 & 163 \\
NGC1637 & $2.05\pm 0.066$& $2.06\pm 0.055$ & 202  & 89552 & 2.62&0.39 & 1.83,1.74 & 116 & 2.56,1.91  & 86 \\
NGC2903 & $1.70\pm 0.023$& $1.77\pm 0.019$ & 642 & 134328 & 5.41&0.30  & 1.62,1.78 & 346 & 1.83,1.95 & 296 \\
NGC3521 & $2.15,\pm 0.035$&$2.25\pm 0.027$ & 737 & 397015 & 6.35&0.30  & 2.08,2.28 & 476 & 2.28,2.40 & 261 \\
NGC3621 & $3.00\pm 0.145$&$2.99\pm 0.131$ & 152  & 283582 & 3.58&0.28  & 2.88,2.10  & 95 & 3.31,2.25 &  57 \\
NGC3627 & $1.55\pm 0.017$&$1.65\pm 0.015$ &879 & 119403 & 4.69&0.28  & 1.48,1.66 & 457 & 1.65,1.83 & 422 \\
NGC5643 & $2.10\pm 0.043$&$2.11\pm 0.049$ & 399 & 179104 & 4.50&0.53  & 1.97,1.82 & 244 & 2.40,2.00 & 155 \\
NGC6300 & $2.31\pm 0.074$&$2.32\pm 0.067$  &  286 & 253731 & 5.43&0.72  &2.04,1.99 & 167 & 3.00,2.20  &119 \\
M 33    & $1.74\pm 0.039$&$1.95\pm 0.030$  &  449   &  7000   &  2.82&0.37 & 1.51,1.83 & 241 & 2.00,2.12 & 208 \\
\hline
\end{tabular}
\tablefoot{PHANGS galaxies observed by \citet{Rosolowsky2021} followed by results from fitting the CO luminosity distribution with the \citet{Maschberger09} method and powerlaw.py algorithm \citep{Alstott2014}. 2nd column is overall slope, where the second number is determined with powerlaw.py. 3rd column is the number of GMCs with a flux greater than the value in col. 4 expressed in units of $\K\kms$ pc$^2$ . Col 4 is designed to be the true completeness limit as explained in the text. Col. 5 gives the median galactocentric radius of the GMCs for that galaxy in kpc and as a fraction of R$_{25}$.  The following two columns provide the inner slope with the two methods and the number of GMCs. The next two columns are the same but for the outer parts. Due to the lower number of clouds, the uncertainties on $\alpha$ are higher when the inner and outer parts are evaluated separately.}
\end{center}
 \label{alph_gmc} 
\end{table*}

\begin{table*}
\caption[]{Fits of mock data}
\begin{center}
\begin{tabular}{lllll}
& Bol. stochastic & Bol. Larson & H$\alpha$ stochastic & H$\alpha$ Larson \\
$\alpha_{cl}$ & $\alpha_{{bol},MK}$, $\alpha_{{bol},py}$ & $\alpha_{{bol},MK}$,$\alpha_{{bol},py}$  & 
$\alpha_{{H\alpha},MK}$,$\alpha_{{H\alpha},py}$  & $\alpha_{{H\alpha},MK}$,$\alpha_{{H\alpha},py}$   \\
\hline
2.3 & 1.90, 1.93 &  2.17, 2.18 & 1.62, 1.66  & 1.93, 1.94  \\
2.0 & 1.82, 1.86 & 1.96, 1.98  & 1.58, 1.62  & 1.79, 1.81  \\
1.7 & 1.75, 1.80 & 1.81, 1.86  & 1.53, 1.58  & 1.69, 1.72  \\
\hline
\end{tabular}
\tablefoot{As in previous tables, the first fit is with the \citet{Maschberger09} method and the second is with the powerlaw.py algorithm \citep{Alstott2014}. First column is the spectral index of the injected GMC or cluster mass distribution.   The following columns provide the spectral indices $\alpha$ of the fits to the bolometric luminosity distribution and the H$\alpha$ luminosity distribution for the stochastic and \citet{Larson1982} IMFs.  This is the data used as input to Fig.~\ref{gmc2hii2fig}. }
\end{center}
 \label{gmc2hii} 
\end{table*}

\section{Conclusions}

The available large high quality samples of HII regions and molecular clouds have been used to look for systematic radial variations of the luminosity function and constraints on the Initial Mass Function (IMF) of stars.  Only galaxies with large numbers of HII regions and/or molecular clouds were treated here to ensure statistical robustness.  

The CO luminosity function always steepens with galactocentric distance.
 The H$\alpha$ luminosity function usually steepens with galactocentric distance.
 As the intersection of the two samples is quite small (3 galaxies), it is difficult to assert that there is a true difference in behavior.
 However, the molecular cloud mass distribution is consistently steeper than that of the HII region H$\alpha$ luminosity distribution
 and this may provide information on the IMF.

The main IMFs proposed in the literature \citep{Salpeter55,Kroupa2001,Chabrier2003} differ in the 
low-mass end and hence only affect luminosities marginally due to the lesser mass stored in low-mass stars in the 
Chabrier and Kroupa IMFs as compared to Salpeter. Therefore the flatter HII region luminosity distribution cannot be due to a change in the low mass IMF. Metallicity differences may also affect the IMF but within this sample the metallicity variation is small. 
It has also been proposed
that massive clouds are required to be able to form the highest mass stars.  Larson (1982, 2003) suggested the maximum stellar mass $M_{*,max} = 0.33 M_{\rm GMC}^{0.43}$.   In this work we compare a fully stochastically sampled IMF with the same IMF with a mass truncation as proposed in \citet{Larson1982} or by the Kroupa group \citep[e.g.][]{Zhou2025}. Allowing for aging is essential as the highest mass stars are short-lived.

Contrary to our initial expectations, the difference between the HII region H$\alpha$ luminosity function and the GMC mass function is consistent with the fully stochastically sampled IMF with no link between $M_{*,max}$ and GMC mass.

\bibliographystyle{aa}
\bibliography{jb}

@ARTICLE{2009A&A...495..479C,
       author = {{Corbelli}, E. and {Verley}, S. and {Elmegreen}, B.~G. and {Giovanardi}, C.},
        title = "{The cluster birthline in M 33}",
      journal = {\aap},
         year = 2009,
        month = feb,
       volume = {495},
       number = {2},
        pages = {479-490},
          doi = {10.1051/0004-6361:200811086},
archivePrefix = {arXiv},
       eprint = {0901.1530},
 primaryClass = {astro-ph.GA},
       adsurl = {https://ui.adsabs.harvard.edu/abs/2009A&A...495..479C}
}

@ARTICLE{Alstott2014,
        AUTHOR             = {Alstott, J. and Bullmore, E. and Plenz, D.},
        JOURNAL            = {PLoS ONE},
        PAGES              = {e85777},
        VOLUME             = {9},
        YEAR               = {2014}
}

@ARTICLE{Braine2018,
   author = {{Braine}, J. and {Rosolowsky}, E. and {Gratier}, P. and {Corbelli}, E. and 
	{Schuster}, K.-F.},
    title = "{Properties and rotation of molecular clouds in M 33}",
  journal = {A\&A},
archivePrefix = "arXiv",
   eprint = {1801.04171},
     year = 2018,
    month = apr,
   volume = 612,
      eid = {A51},
    pages = {A51},
      doi = {10.1051/0004-6361/201732405},
   adsurl = {http://adsabs.harvard.edu/abs/2018A%26A...612A..51B},
}

@ARTICLE{Calzetti2012,
       author = {{Calzetti}, D. and {Liu}, G. and {Koda}, J.},
        title = "{Star Formation Laws: The Effects of Gas Cloud Sampling}",
      journal = {ApJ},
         year = 2012,
        month = jun,
       volume = {752},
       number = {2},
          eid = {98},
        pages = {98},
          doi = {10.1088/0004-637X/752/2/98},
archivePrefix = {arXiv},
       eprint = {1204.5659},
 primaryClass = {astro-ph.GA},
       adsurl = {https://ui.adsabs.harvard.edu/abs/2012ApJ...752...98C},
}

@ARTICLE{Chabrier2003,
       author = {{Chabrier}, Gilles},
        title = "{Galactic Stellar and Substellar Initial Mass Function}",
      journal = {PASP},
         year = 2003,
        month = jul,
       volume = {115},
       number = {809},
        pages = {763-795},
          doi = {10.1086/376392},
archivePrefix = {arXiv},
       eprint = {astro-ph/0304382},
 primaryClass = {astro-ph},
       adsurl = {https://ui.adsabs.harvard.edu/abs/2003PASP..115..763C},
}

@ARTICLE{Chevance2020,
       author = {{Chevance}, M{\'e}lanie and {Kruijssen}, J.~M. Diederik and {Hygate}, Alexander P.~S. and {Schruba}, Andreas and {Longmore}, Steven N. and {Groves}, Brent and {Henshaw}, Jonathan D. and {Herrera}, Cinthya N. and {Hughes}, Annie and {Jeffreson}, Sarah M.~R. and {Lang}, Philipp and {Leroy}, Adam K. and {Meidt}, Sharon E. and {Pety}, J{\'e}r{\^o}me and {Razza}, Alessandro and {Rosolowsky}, Erik and {Schinnerer}, Eva and {Bigiel}, Frank and {Blanc}, Guillermo A. and {Emsellem}, Eric and {Faesi}, Christopher M. and {Glover}, Simon C.~O. and {Haydon}, Daniel T. and {Ho}, I.-Ting and {Kreckel}, Kathryn and {Lee}, Janice C. and {Liu}, Daizhong and {Querejeta}, Miguel and {Saito}, Toshiki and {Sun}, Jiayi and {Usero}, Antonio and {Utomo}, Dyas},
        title = "{The lifecycle of molecular clouds in nearby star-forming disc galaxies}",
      journal = {MNRAS},
         year = 2020,
        month = apr,
       volume = {493},
       number = {2},
        pages = {2872-2909},
          doi = {10.1093/mnras/stz3525},
archivePrefix = {arXiv},
       eprint = {1911.03479},
 primaryClass = {astro-ph.GA},
       adsurl = {https://ui.adsabs.harvard.edu/abs/2020MNRAS.493.2872C},
}

@ARTICLE{Colombo2014a,
   author = {{Colombo}, D. and {Hughes}, A. and {Schinnerer}, E. and {Meidt}, S.~E. and 
	{Leroy}, A.~K. and {Pety}, J. and {Dobbs}, C.~L. and {Garc{\'{\i}}a-Burillo}, S. and 
	{Dumas}, G. and {Thompson}, T.~A. and {Schuster}, K.~F. and 
	{Kramer}, C.},
    title = "{The PdBI Arcsecond Whirlpool Survey (PAWS): Environmental Dependence of Giant Molecular Cloud Properties in M51}",
  journal = {ApJ},
archivePrefix = "arXiv",
   eprint = {1401.1505},
 primaryClass = "astro-ph.GA",
     year = 2014,
    month = mar,
   volume = 784,
      eid = {3},
    pages = {3},
      doi = {10.1088/0004-637X/784/1/3},
   adsurl = {http://adsabs.harvard.edu/abs/2014ApJ...784....3C},
}

@ARTICLE{Corbelli17,
   author = {{Corbelli}, E. and {Braine}, J. and {Bandiera}, R. and {Brouillet}, N. and 
	{Combes}, F. and {Druard}, C. and {Gratier}, P. and {Mata}, J. and 
	{Schuster}, K. and {Xilouris}, M. and {Palla}, F.},
    title = "{From molecules to young stellar clusters: the star formation cycle across the disk of M 33}",
  journal = {A\&A},
archivePrefix = "arXiv",
   eprint = {1703.09183},
     year = 2017,
    month = may,
   volume = 601,
      eid = {A146},
    pages = {A146},
      doi = {10.1051/0004-6361/201630034},
   adsurl = {http://adsabs.harvard.edu/abs/2017A%26A...601A.146C},
}

@ARTICLE{daSilva2014,
       author = {{da Silva}, Robert L. and {Fumagalli}, Michele and {Krumholz}, Mark R.},
        title = "{SLUG - Stochastically Lighting Up Galaxies - II. Quantifying the effects of stochasticity on star formation rate indicators}",
      journal = {MNRAS},
         year = 2014,
        month = nov,
       volume = {444},
       number = {4},
        pages = {3275-3287},
          doi = {10.1093/mnras/stu1688},
archivePrefix = {arXiv},
       eprint = {1403.4605},
 primaryClass = {astro-ph.GA},
       adsurl = {https://ui.adsabs.harvard.edu/abs/2014MNRAS.444.3275D},
}

@ARTICLE{Demachi2024,
       author = {{Demachi}, Fumika and {Fukui}, Yasuo and {Yamada}, Rin I. and {Tachihara}, Kengo and {Hayakawa}, Takahiro and {Tokuda}, Kazuki and {Fujita}, Shinji and {Kobayashi}, Masato I.~N. and {Muraoka}, Kazuyuki and {Konishi}, Ayu and {Tsuge}, Kisetsu and {Onishi}, Toshikazu and {Kawamura}, Akiko},
        title = "{Giant molecular clouds and their type classification in M 74: Toward understanding star formation and cloud evolution}",
      journal = {PASJ},
         year = 2024,
        month = oct,
       volume = {76},
       number = {5},
        pages = {1059-1083},
          doi = {10.1093/pasj/psae071},
archivePrefix = {arXiv},
       eprint = {2305.19192},
 primaryClass = {astro-ph.GA},
       adsurl = {https://ui.adsabs.harvard.edu/abs/2024PASJ...76.1059D},
}

@ARTICLE{Dickman86,
    author = "{Dickman}, R. L. and {Snell}, Ronald L. and {Schloerb}, F. Peter",
    title = "Carbon monoxide as an extragalactic mass tracer",
    journal = {ApJ},
    year = 1986,
    month = oct,
    volume = 309,
    pages = "326--330",
}

@ARTICLE{Druard14,
   author = {{Druard}, C. and {Braine}, J. and {Schuster}, K.~F. and {Schneider}, N. and 
	{Gratier}, P. and {Bontemps}, S. and {Boquien}, M. and {Combes}, F. and 
	{Corbelli}, E. and {Henkel}, C. and {Herpin}, F. and {Kramer}, C. and 
	{van der Tak}, F. and {van der Werf}, P.},
    title = "{The IRAM M 33 CO(2-1) survey. A complete census of molecular gas out to 7 kpc}",
  journal = {A\&A},
archivePrefix = "arXiv",
   eprint = {1405.5884},
     year = 2014,
    month = jul,
   volume = 567,
      eid = {A118},
    pages = {A118},
      doi = {10.1051/0004-6361/201423682},
   adsurl = {http://adsabs.harvard.edu/abs/2014A%26A...567A.118D},
}

@ARTICLE{Eker2015,
       author = {{Eker}, Z. and {Soydugan}, F. and {Soydugan}, E. and {Bilir}, S. and {Yaz G{\"o}k{\c{c}}e}, E. and {Steer}, I. and {T{\"u}ys{\"u}z}, M. and {{\c{S}}eny{\"u}z}, T. and {Demircan}, O.},
        title = "{Main-Sequence Effective Temperatures from a Revised Mass-Luminosity Relation Based on Accurate Properties}",
      journal = {AJ},
         year = 2015,
        month = apr,
       volume = {149},
       number = {4},
          eid = {131},
        pages = {131},
          doi = {10.1088/0004-6256/149/4/131},
archivePrefix = {arXiv},
       eprint = {1501.06585},
 primaryClass = {astro-ph.SR},
       adsurl = {https://ui.adsabs.harvard.edu/abs/2015AJ....149..131E},
}

@ARTICLE{Eker2018,
       author = {{Eker}, Z. and {Bak{\i}{\c{s}}}, V. and {Bilir}, S. and {Soydugan}, F. and {Steer}, I. and {Soydugan}, E. and {Bak{\i}{\c{s}}}, H. and {Ali{\c{c}}avu{\c{s}}}, F. and {Aslan}, G. and {Alpsoy}, M.},
        title = "{Interrelated main-sequence mass-luminosity, mass-radius, and mass-effective temperature relations}",
      journal = {MNRAS},
         year = 2018,
        month = oct,
       volume = {479},
       number = {4},
        pages = {5491-5511},
          doi = {10.1093/mnras/sty1834},
archivePrefix = {arXiv},
       eprint = {1807.02568},
 primaryClass = {astro-ph.SR},
       adsurl = {https://ui.adsabs.harvard.edu/abs/2018MNRAS.479.5491E},
}

@ARTICLE{Evans2009,
       author = {{Evans}, Neal J., II and {Dunham}, Michael M. and {J{\o}rgensen}, Jes K. and {Enoch}, Melissa L. and {Mer{\'\i}n}, Bruno and {van Dishoeck}, Ewine F. and {Alcal{\'a}}, Juan M. and {Myers}, Philip C. and {Stapelfeldt}, Karl R. and {Huard}, Tracy L. and {Allen}, Lori E. and {Harvey}, Paul M. and {van Kempen}, Tim and {Blake}, Geoffrey A. and {Koerner}, David W. and {Mundy}, Lee G. and {Padgett}, Deborah L. and {Sargent}, Anneila I.},
        title = "{The Spitzer c2d Legacy Results: Star-Formation Rates and Efficiencies; Evolution and Lifetimes}",
      journal = {ApJD},
         year = 2009,
        month = apr,
       volume = {181},
       number = {2},
        pages = {321-350},
          doi = {10.1088/0067-0049/181/2/321},
archivePrefix = {arXiv},
       eprint = {0811.1059},
 primaryClass = {astro-ph},
       adsurl = {https://ui.adsabs.harvard.edu/abs/2009ApJS..181..321E},
}

@ARTICLE{Gratier12,
    author = {{Gratier}, P. and {Braine}, J. and {Rodriguez-Fernandez}, N.~J. and 
	{Schuster}, K.~F. and {Kramer}, C. and {Corbelli}, E. and {Combes}, F. and 
	{Brouillet}, N. and {van der Werf}, P.~P. and {R{\"o}llig}, M.},
    title = "{Giant molecular clouds in the Local Group galaxy M 33}",
  journal = {A\&A},
archivePrefix = "arXiv",
   eprint = {1111.4320},
 primaryClass = "astro-ph.CO",
     year = 2012,
    month = jun,
   volume = 542,
      eid = {108},
    pages = {108},
      doi = {10.1051/0004-6361/201116612},
   adsurl = {http://adsabs.harvard.edu/abs/2012A%26A...542A.108G},
}

@ARTICLE{Hopkins2018,
       author = {{Hopkins}, A.~M.},
        title = "{The Dawes Review 8: Measuring the Stellar Initial Mass Function}",
      journal = {PASA},
         year = 2018,
        month = nov,
       volume = {35},
          eid = {e039},
        pages = {e039},
          doi = {10.1017/pasa.2018.29},
archivePrefix = {arXiv},
       eprint = {1807.09949},
 primaryClass = {astro-ph.GA},
       adsurl = {https://ui.adsabs.harvard.edu/abs/2018PASA...35...39H},
}

@ARTICLE{Kobayashi2017,
       author = {{Kobayashi}, Masato I.~N. and {Inutsuka}, Shu-ichiro and {Kobayashi}, Hiroshi and {Hasegawa}, Kenji},
        title = "{Evolutionary Description of Giant Molecular Cloud Mass Functions on Galactic Disks}",
      journal = {ApJ},
         year = 2017,
        month = feb,
       volume = {836},
       number = {2},
          eid = {175},
        pages = {175},
          doi = {10.3847/1538-4357/836/2/175},
archivePrefix = {arXiv},
       eprint = {1701.03781},
 primaryClass = {astro-ph.GA},
       adsurl = {https://ui.adsabs.harvard.edu/abs/2017ApJ...836..175K},
}

@ARTICLE{Kroupa2001,
       author = {{Kroupa}, Pavel},
        title = "{On the variation of the initial mass function}",
      journal = {MNRAS},
         year = 2001,
        month = apr,
       volume = {322},
       number = {2},
        pages = {231-246},
          doi = {10.1046/j.1365-8711.2001.04022.x},
archivePrefix = {arXiv},
       eprint = {astro-ph/0009005},
 primaryClass = {astro-ph},
       adsurl = {https://ui.adsabs.harvard.edu/abs/2001MNRAS.322..231K},
}

@ARTICLE{Larson1982,
       author = {{Larson}, R.~B.},
        title = "{Mass spectra of young stars.}",
      journal = {MNRAS},
         year = 1982,
        month = jul,
       volume = {200},
        pages = {159-174},
          doi = {10.1093/mnras/200.2.159},
       adsurl = {https://ui.adsabs.harvard.edu/abs/1982MNRAS.200..159L},
}

@ARTICLE{Larson1998,
       author = {{Larson}, Richard B.},
        title = "{Early star formation and the evolution of the stellar initial mass function in galaxies}",
      journal = {MNRAS},
         year = 1998,
        month = dec,
       volume = {301},
       number = {2},
        pages = {569-581},
          doi = {10.1046/j.1365-8711.1998.02045.x},
archivePrefix = {arXiv},
       eprint = {astro-ph/9808145},
 primaryClass = {astro-ph},
       adsurl = {https://ui.adsabs.harvard.edu/abs/1998MNRAS.301..569L},
}

@INPROCEEDINGS{Larson2003,
       author = {{Larson}, R.~B.},
        title = "{The Stellar Initial Mass Function and Beyond (Invited Review)}",
    booktitle = {Galactic Star Formation Across the Stellar Mass Spectrum},
         year = 2003,
       editor = {{De Buizer}, James M. and {van der Bliek}, Nicole S.},
       series = {Astronomical Society of the Pacific Conference Series},
       volume = {287},
        month = jan,
        pages = {65-80},
          doi = {10.48550/arXiv.astro-ph/0205466},
archivePrefix = {arXiv},
       eprint = {astro-ph/0205466},
 primaryClass = {astro-ph},
       adsurl = {https://ui.adsabs.harvard.edu/abs/2003ASPC..287...65L},
}

@ARTICLE{Leroy2013,
       author = {{Leroy}, Adam K. and {Walter}, Fabian and {Sandstrom}, Karin and {Schruba}, Andreas and {Munoz-Mateos}, Juan-Carlos and {Bigiel}, Frank and {Bolatto}, Alberto and {Brinks}, Elias and {de Blok}, W.~J.~G. and {Meidt}, Sharon and {Rix}, Hans-Walter and {Rosolowsky}, Erik and {Schinnerer}, Eva and {Schuster}, Karl-Friedrich and {Usero}, Antonio},
        title = "{Molecular Gas and Star Formation in nearby Disk Galaxies}",
      journal = {AJ},
         year = 2013,
        month = aug,
       volume = {146},
       number = {2},
          eid = {19},
        pages = {19},
          doi = {10.1088/0004-6256/146/2/19},
archivePrefix = {arXiv},
       eprint = {1301.2328},
 primaryClass = {astro-ph.CO},
       adsurl = {https://ui.adsabs.harvard.edu/abs/2013AJ....146...19L},
}

@ARTICLE{Leroy2021,
       author = {{Leroy}, Adam K. and {Schinnerer}, Eva and {Hughes}, Annie and {Rosolowsky}, Erik and {Pety}, J{\'e}r{\^o}me and {Schruba}, Andreas and {Usero}, Antonio and {Blanc}, Guillermo A. and {Chevance}, M{\'e}lanie and {Emsellem}, Eric and {Faesi}, Christopher M. and {Herrera}, Cinthya N. and {Liu}, Daizhong and {Meidt}, Sharon E. and {Querejeta}, Miguel and {Saito}, Toshiki and {Sandstrom}, Karin M. and {Sun}, Jiayi and {Williams}, Thomas G. and {Anand}, Gagandeep S. and {Barnes}, Ashley T. and {Behrens}, Erica A. and {Belfiore}, Francesco and {Benincasa}, Samantha M. and {Be{\v{s}}li{\'c}}, Ivana and {Bigiel}, Frank and {Bolatto}, Alberto D. and {den Brok}, Jakob S. and {Cao}, Yixian and {Chandar}, Rupali and {Chastenet}, J{\'e}r{\'e}my and {Chiang}, I-Da and {Congiu}, Enrico and {Dale}, Daniel A. and {Deger}, Sinan and {Eibensteiner}, Cosima and {Egorov}, Oleg V. and {Garc{\'\i}a-Rodr{\'\i}guez}, Axel and {Glover}, Simon C.~O. and {Grasha}, Kathryn and {Henshaw}, Jonathan D. and {Ho}, I. -Ting and {Kepley}, Amanda A. and {Kim}, Jaeyeon and {Klessen}, Ralf S. and {Kreckel}, Kathryn and {Koch}, Eric W. and {Kruijssen}, J.~M. Diederik and {Larson}, Kirsten L. and {Lee}, Janice C. and {Lopez}, Laura A. and {Machado}, Josh and {Mayker}, Ness and {McElroy}, Rebecca and {Murphy}, Eric J. and {Ostriker}, Eve C. and {Pan}, Hsi-An and {Pessa}, Ismael and {Puschnig}, Johannes and {Razza}, Alessandro and {S{\'a}nchez-Bl{\'a}zquez}, Patricia and {Santoro}, Francesco and {Sardone}, Amy and {Scheuermann}, Fabian and {Sliwa}, Kazimierz and {Sormani}, Mattia C. and {Stuber}, Sophia K. and {Thilker}, David A. and {Turner}, Jordan A. and {Utomo}, Dyas and {Watkins}, Elizabeth J. and {Whitmore}, Bradley},
        title = "{PHANGS-ALMA: Arcsecond CO(2-1) Imaging of Nearby Star-forming Galaxies}",
      journal = {ApJS},
         year = 2021,
        month = dec,
       volume = {257},
       number = {2},
          eid = {43},
        pages = {43},
          doi = {10.3847/1538-4365/ac17f3},
archivePrefix = {arXiv},
       eprint = {2104.07739},
 primaryClass = {astro-ph.GA},
       adsurl = {https://ui.adsabs.harvard.edu/abs/2021ApJS..257...43L}
}

@ARTICLE{Leroy2025,
       author = {{Leroy}, Adam K. and {Sun}, Jiayi and {Meidt}, Sharon and {Agertz}, Oscar and {Chiang}, I.-Da and {Gensior}, Jindra and {Glover}, Simon C.~O. and {Gnedin}, Oleg Y. and {Hughes}, Annie and {Schinnerer}, Eva and {Barnes}, Ashley T. and {Bigiel}, Frank and {Bolatto}, Alberto D. and {Colombo}, Dario and {den Brok}, Jakob and {Chevance}, M{\'e}lanie and {Chown}, Ryan and {Eibensteiner}, Cosima and {Gleis}, Damian R. and {Grasha}, Kathryn and {Henshaw}, Jonathan D. and {Klessen}, Ralf S. and {Koch}, Eric W. and {Oakes}, Elias K. and {Pan}, Hsi-An and {Querejeta}, Miguel and {Rosolowsky}, Erik and {Saito}, Toshiki and {Sandstrom}, Karin and {Sarbadhicary}, Sumit K. and {Teng}, Yu-Hsuan and {Usero}, Antonio and {Utomo}, Dyas and {Williams}, Thomas G.},
        title = "{Cloud-scale Gas Properties, Depletion Times, and Star Formation Efficiency per Freefall Time in PHANGS{\textendash}ALMA}",
      journal = {ApJ},
         year = 2025,
        month = may,
       volume = {985},
       number = {1},
          eid = {14},
        pages = {14},
          doi = {10.3847/1538-4357/adbcab},
archivePrefix = {arXiv},
       eprint = {2502.04481},
 primaryClass = {astro-ph.GA},
       adsurl = {https://ui.adsabs.harvard.edu/abs/2025ApJ...985...14L},
}

@ARTICLE{Lin2017,
       author = {{Lin}, Zesen and {Hu}, Ning and {Kong}, Xu and {Gao}, Yulong and {Zou}, Hu and {Wang}, Enci and {Cheng}, Fuzhen and {Fang}, Guanwen and {Lin}, Lin and {Wang}, Jing},
        title = "{Spectroscopic Observation and Analysis of H II Regions in M33 with MMT: Temperatures and Oxygen Abundances}",
      journal = {ApJ},
         year = 2017,
        month = jun,
       volume = {842},
       number = {2},
          eid = {97},
        pages = {97},
          doi = {10.3847/1538-4357/aa6f14},
archivePrefix = {arXiv},
       eprint = {1704.06935},
 primaryClass = {astro-ph.GA},
       adsurl = {https://ui.adsabs.harvard.edu/abs/2017ApJ...842...97L},
}

@ARTICLE{Maschberger09,
   author = {{Maschberger}, T. and {Kroupa}, P.},
    title = "{Estimators for the exponent and upper limit, and goodness-of-fit tests for (truncated) power-law distributions}",
  journal = {MNRAS},
archivePrefix = "arXiv",
   eprint = {0905.0474},
 primaryClass = "astro-ph.IM",
     year = 2009,
    month = may,
   volume = 395,
    pages = {931-942},
      doi = {10.1111/j.1365-2966.2009.14577.x},
   adsurl = {http://adsabs.harvard.edu/abs/2009MNRAS.395..931M},
}

@ARTICLE{Murgia02,
   author = {{Murgia}, M. and {Crapsi}, A. and {Moscadelli}, L. and {Gregorini}, L.},
    title = "{Radio continuum and CO emission in star-forming galaxies}",
  journal = {A\&A},
   eprint = {arXiv:astro-ph/0201478},
     year = 2002,
    month = apr,
   volume = 385,
    pages = {412-424},
}

@ARTICLE{Murray2011,
       author = {{Murray}, Norman},
        title = "{Star Formation Efficiencies and Lifetimes of Giant Molecular Clouds in the Milky Way}",
      journal = {ApJ},
         year = 2011,
        month = mar,
       volume = {729},
       number = {2},
          eid = {133},
        pages = {133},
          doi = {10.1088/0004-637X/729/2/133},
archivePrefix = {arXiv},
       eprint = {1007.3270},
 primaryClass = {astro-ph.GA},
       adsurl = {https://ui.adsabs.harvard.edu/abs/2011ApJ...729..133M},
}

@ARTICLE{Rosolowsky05,
   author = {{Rosolowsky}, E.},
    title = "{The Mass Spectra of Giant Molecular Clouds in the Local Group}",
  journal = {PASP},
   eprint = {arXiv:astro-ph/0508679},
     year = 2005,
    month = dec,
   volume = 117,
    pages = {1403-1410},
      doi = {10.1086/497582},
   adsurl = {http://adsabs.harvard.edu/abs/2005PASP..117.1403R},
}

@ARTICLE{Rosolowsky07,
   author = {{Rosolowsky}, E. and {Keto}, E. and {Matsushita}, S. and {Willner}, S.~P.},
    title = "{High-Resolution Molecular Gas Maps of M33}",
  journal = {ApJ},
   eprint = {arXiv:astro-ph/0703006},
     year = 2007,
    month = jun,
   volume = 661,
    pages = {830-844},
}

@ARTICLE{Rosolowsky2021,
       author = {{Rosolowsky}, Erik and {Hughes}, Annie and {Leroy}, Adam K. and {Sun}, Jiayi and {Querejeta}, Miguel and {Schruba}, Andreas and {Usero}, Antonio and {Herrera}, Cinthya N. and {Liu}, Daizhong and {Pety}, J{\'e}r{\^o}me and {Saito}, Toshiki and {Be{\v{s}}li{\'c}}, Ivana and {Bigiel}, Frank and {Blanc}, Guillermo and {Chevance}, M{\'e}lanie and {Dale}, Daniel A. and {Deger}, Sinan and {Faesi}, Christopher M. and {Glover}, Simon C.~O. and {Henshaw}, Jonathan D. and {Klessen}, Ralf S. and {Kruijssen}, J.~M. Diederik and {Larson}, Kirsten and {Lee}, Janice and {Meidt}, Sharon and {Mok}, Angus and {Schinnerer}, Eva and {Thilker}, David A. and {Williams}, Thomas G.},
        title = "{Giant molecular cloud catalogues for PHANGS-ALMA: methods and initial results}",
      journal = {MNRAS},
         year = 2021,
        month = mar,
       volume = {502},
       number = {1},
        pages = {1218-1245},
          doi = {10.1093/mnras/stab085},
archivePrefix = {arXiv},
       eprint = {2101.04697},
 primaryClass = {astro-ph.GA},
       adsurl = {https://ui.adsabs.harvard.edu/abs/2021MNRAS.502.1218R},
}

@ARTICLE{Salpeter55,
    author = {{Salpeter}, E.\ E.},
    title = "The Luminosity Function and Stellar Evolution.",
    journal = {ApJ},
    year = 1955,
    month = jan,
    volume = 121,
    pages = {161},
}

@ARTICLE{Santoro2022,
       author = {{Santoro}, Francesco and {Kreckel}, Kathryn and {Belfiore}, Francesco and {Groves}, Brent and {Congiu}, Enrico and {Thilker}, David A. and {Blanc}, Guillermo A. and {Schinnerer}, Eva and {Ho}, I. -Ting and {Kruijssen}, J.~M. Diederik and {Meidt}, Sharon and {Klessen}, Ralf S. and {Schruba}, Andreas and {Querejeta}, Miguel and {Pessa}, Ismael and {Chevance}, M{\'e}lanie and {Kim}, Jaeyeon and {Emsellem}, Eric and {McElroy}, Rebecca and {Barnes}, Ashley T. and {Bigiel}, Frank and {Boquien}, M{\'e}d{\'e}ric and {Dale}, Daniel A. and {Glover}, Simon C.~O. and {Grasha}, Kathryn and {Lee}, Janice and {Leroy}, Adam K. and {Pan}, Hsi-An and {Rosolowsky}, Erik and {Saito}, Toshiki and {Sanchez-Blazquez}, Patricia and {Watkins}, Elizabeth J. and {Williams}, Thomas G.},
        title = "{PHANGS-MUSE: The H II region luminosity function of local star-forming galaxies}",
      journal = {A\&A},
         year = 2022,
        month = feb,
       volume = {658},
          eid = {A188},
        pages = {A188},
          doi = {10.1051/0004-6361/202141907},
archivePrefix = {arXiv},
       eprint = {2111.09362},
 primaryClass = {astro-ph.GA},
       adsurl = {https://ui.adsabs.harvard.edu/abs/2022A&A...658A.188S},
}

@ARTICLE{Schinnerer2019,
       author = {{Schinnerer}, Eva and {Hughes}, Annie and {Leroy}, Adam and {Groves}, Brent and {Blanc}, Guillermo A. and {Kreckel}, Kathryn and {Bigiel}, Frank and {Chevance}, M{\'e}lanie and {Dale}, Daniel and {Emsellem}, Eric and {Faesi}, Christopher and {Glover}, Simon and {Grasha}, Kathryn and {Henshaw}, Jonathan and {Hygate}, Alexander and {Kruijssen}, J.~M. Diederik and {Meidt}, Sharon and {Pety}, Jerome and {Querejeta}, Miguel and {Rosolowsky}, Erik and {Saito}, Toshiki and {Schruba}, Andreas and {Sun}, Jiayi and {Utomo}, Dyas},
        title = "{The Gas-Star Formation Cycle in Nearby Star-forming Galaxies. I. Assessment of Multi-scale Variations}",
      journal = {ApJ},
         year = 2019,
        month = dec,
       volume = {887},
       number = {1},
          eid = {49},
        pages = {49},
          doi = {10.3847/1538-4357/ab50c2},
archivePrefix = {arXiv},
       eprint = {1910.10520},
 primaryClass = {astro-ph.GA},
       adsurl = {https://ui.adsabs.harvard.edu/abs/2019ApJ...887...49S},
}

@ARTICLE{Schinnerer2024,
       author = {{Schinnerer}, E. and {Leroy}, A.~K.},
        title = "{Molecular Gas and the Star-Formation Process on Cloud Scales in Nearby Galaxies}",
      journal = {ARA\&A},
         year = 2024,
        month = sep,
       volume = {62},
       number = {1},
        pages = {369-436},
          doi = {10.1146/annurev-astro-071221-052651},
archivePrefix = {arXiv},
       eprint = {2403.19843},
 primaryClass = {astro-ph.GA},
       adsurl = {https://ui.adsabs.harvard.edu/abs/2024ARA&A..62..369S},
}

@ARTICLE{Solomon87,
   author = {{Solomon}, P.~M. and {Rivolo}, A.~R. and {Barrett}, J. and {Yahil}, A.},
    title = "{Mass, luminosity, and line width relations of Galactic molecular clouds}",
  journal = {ApJ},
     year = 1987,
    month = aug,
   volume = 319,
    pages = {730-741},
      doi = {10.1086/165493},
}

@ARTICLE{Sternberg2003,
       author = {{Sternberg}, Amiel and {Hoffmann}, Tadziu L. and {Pauldrach}, A.~W.~A.},
        title = "{Ionizing Photon Emission Rates from O- and Early B-Type Stars and Clusters}",
      journal = {ApJ},
         year = 2003,
        month = dec,
       volume = {599},
       number = {2},
        pages = {1333-1343},
          doi = {10.1086/379506},
archivePrefix = {arXiv},
       eprint = {astro-ph/0312232},
 primaryClass = {astro-ph},
       adsurl = {https://ui.adsabs.harvard.edu/abs/2003ApJ...599.1333S},
}

@ARTICLE{Teng2024,
       author = {{Teng}, Yu-Hsuan and {Chiang}, I-Da and {Sandstrom}, Karin M. and {Sun}, Jiayi and {Leroy}, Adam K. and {Bolatto}, Alberto D. and {Usero}, Antonio and {Ostriker}, Eve C. and {Querejeta}, Miguel and {Chastenet}, J{\'e}r{\'e}my and {Bigiel}, Frank and {Boquien}, M{\'e}d{\'e}ric and {den Brok}, Jakob and {Cao}, Yixian and {Chevance}, M{\'e}lanie and {Chown}, Ryan and {Colombo}, Dario and {Eibensteiner}, Cosima and {Glover}, Simon C.~O. and {Grasha}, Kathryn and {Henshaw}, Jonathan D. and {Jim{\'e}nez-Donaire}, Mar{\'\i}a J. and {Liu}, Daizhong and {Murphy}, Eric J. and {Pan}, Hsi-An and {Stuber}, Sophia K. and {Williams}, Thomas G.},
        title = "{Star Formation Efficiency in Nearby Galaxies Revealed with a New CO-to-H$_{2}$ Conversion Factor Prescription}",
      journal = {ApJ},
         year = 2024,
        month = jan,
       volume = {961},
       number = {1},
          eid = {42},
        pages = {42},
          doi = {10.3847/1538-4357/ad10ae},
archivePrefix = {arXiv},
       eprint = {2310.16037},
 primaryClass = {astro-ph.GA},
       adsurl = {https://ui.adsabs.harvard.edu/abs/2024ApJ...961...42T},
}

@ARTICLE{Weidner2004,
       author = {{Weidner}, C. and {Kroupa}, P.},
        title = "{Evidence for a fundamental stellar upper mass limit from clustered star formation}",
      journal = {MNRAS},
         year = 2004,
        month = feb,
       volume = {348},
       number = {1},
        pages = {187-191},
          doi = {10.1111/j.1365-2966.2004.07340.x},
archivePrefix = {arXiv},
       eprint = {astro-ph/0310860},
 primaryClass = {astro-ph},
       adsurl = {https://ui.adsabs.harvard.edu/abs/2004MNRAS.348..187W},
}

@ARTICLE{Weidner2006,
       author = {{Weidner}, Carsten and {Kroupa}, Pavel},
        title = "{The maximum stellar mass, star-cluster formation and composite stellar populations}",
      journal = {MNRAS},
         year = 2006,
        month = feb,
       volume = {365},
       number = {4},
        pages = {1333-1347},
          doi = {10.1111/j.1365-2966.2005.09824.x},
archivePrefix = {arXiv},
       eprint = {astro-ph/0511331},
 primaryClass = {astro-ph},
       adsurl = {https://ui.adsabs.harvard.edu/abs/2006MNRAS.365.1333W},
}

@ARTICLE{Zhou2025,
       author = {{Zhou}, J.~W. and {Kroupa}, Pavel and {Dib}, Sami},
        title = "{The most massive star clusters in molecular clouds: insights from the integrated cloud-wide initial mass function (ICIMF) theory}",
      journal = {MNRAS},
         year = 2025,
        month = aug,
       volume = {541},
       number = {2},
        pages = {1276-1285},
          doi = {10.1093/mnras/staf1070},
archivePrefix = {arXiv},
       eprint = {2506.23096},
 primaryClass = {astro-ph.GA},
       adsurl = {https://ui.adsabs.harvard.edu/abs/2025MNRAS.541.1276Z},
}

\begin{appendix}
\nolinenumbers

\section{Further characterization of the IMF and stellar properties}

The figures in this section illustrate stellar and star cluster properties.  While our baseline is a Chabrier IMF from 0.08--100\msun\ and the Eker-Sternberg Mass-Luminosity relation, the figures show the differences.  Fig.~\ref{rad-lum} shows how the luminosity, radius, effective temperature, and number of ionizing photons vary with mass for the different mass-luminosity relations and a Chabrier IMF from 0.08 to 100~\msun.  

Fig.~\ref{randomness} shows 1000 simulated clusters, each with 300\msun\ of stars, to illustrate the importance of the most massive star in a cluster and how that varies when randomly sampling the IMF.  The top panel shows the distribution of stellar masses for the 3 IMFs examined and how the median mass of the most massive star varies.  The lower panel shows the total cluster luminosity for the 3 IMFs as a function of the most massive star in the cluster ($M_{\rm cluster} = 300$\msun) and the solid curve shows the luminosity of that star.  It is apparent that the most massive star generates the majority of the cluster luminosity in most cases when its mass is $M_{*,max} > 15$\msun. 

Fig.~\ref{starlife} shows how clusters age as a function of time and  choice of IMF.
The top panel shows the average cluster mass fraction remaining as a function of time and (right hand scale) the stellar lifetime, shown as the mass of the most massive star in a cluster as a function of time.
The lower panel displays how the mean light-to-mass ratio of a cluster varies with time and how the number of ionizing photons (right hand scale) decreases with cluster age.

\begin{figure}
        \centering
        \includegraphics[width=\hsize{}]{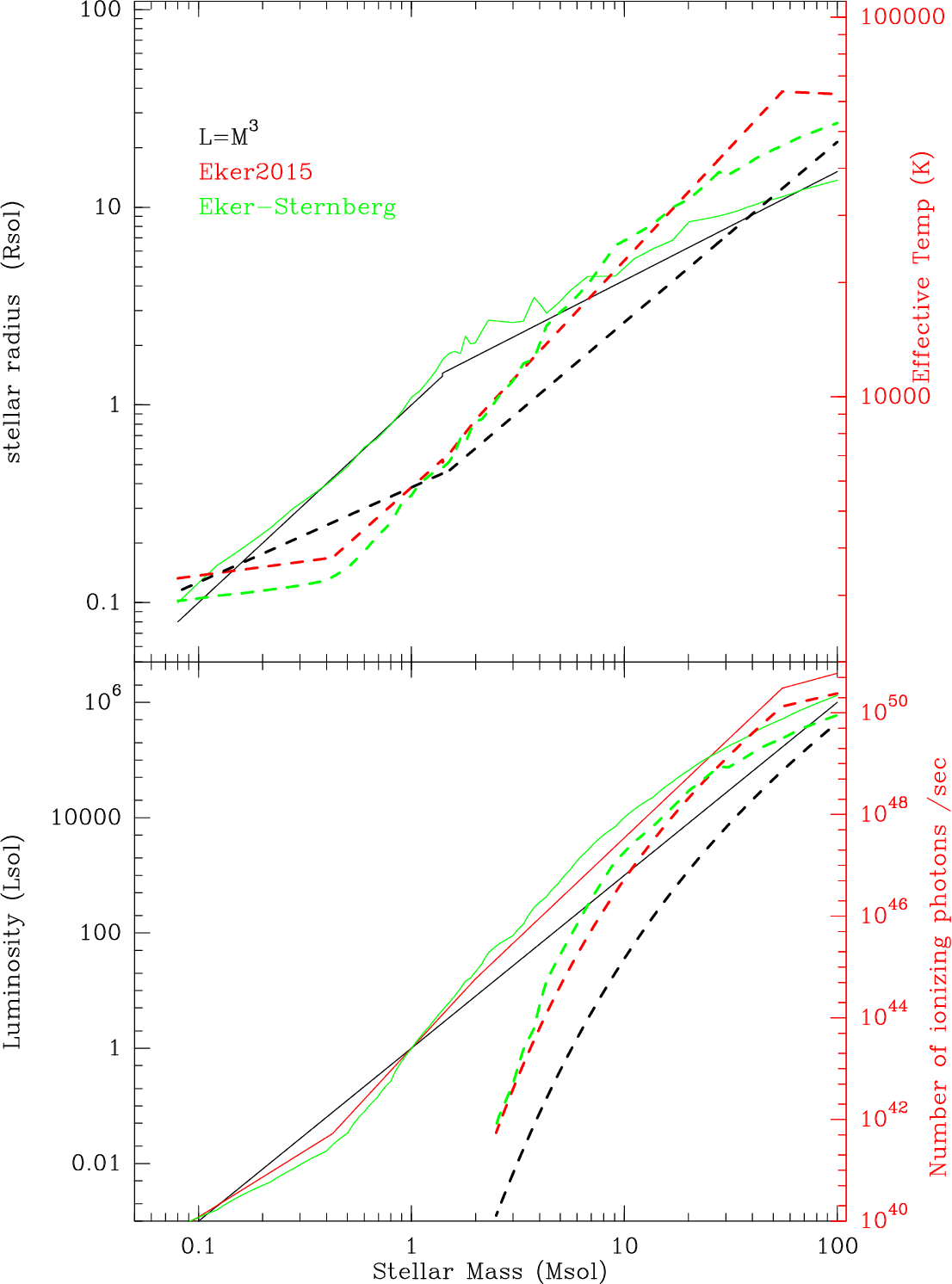}
        \caption{Mass-Luminosity used in this work. Black line is the simple $L=M^3$ relation, red is the generalized \citet{Eker2015}, and green the \citet{Eker2018} and \citet{Sternberg2003} MLR.  The dashed  lines show the property on the right-hand y-axis (Temperature and number of ionizing photons).}
        \label{rad-lum} 
\end{figure}

\begin{figure}
        \centering
        \includegraphics[width=\hsize{}]{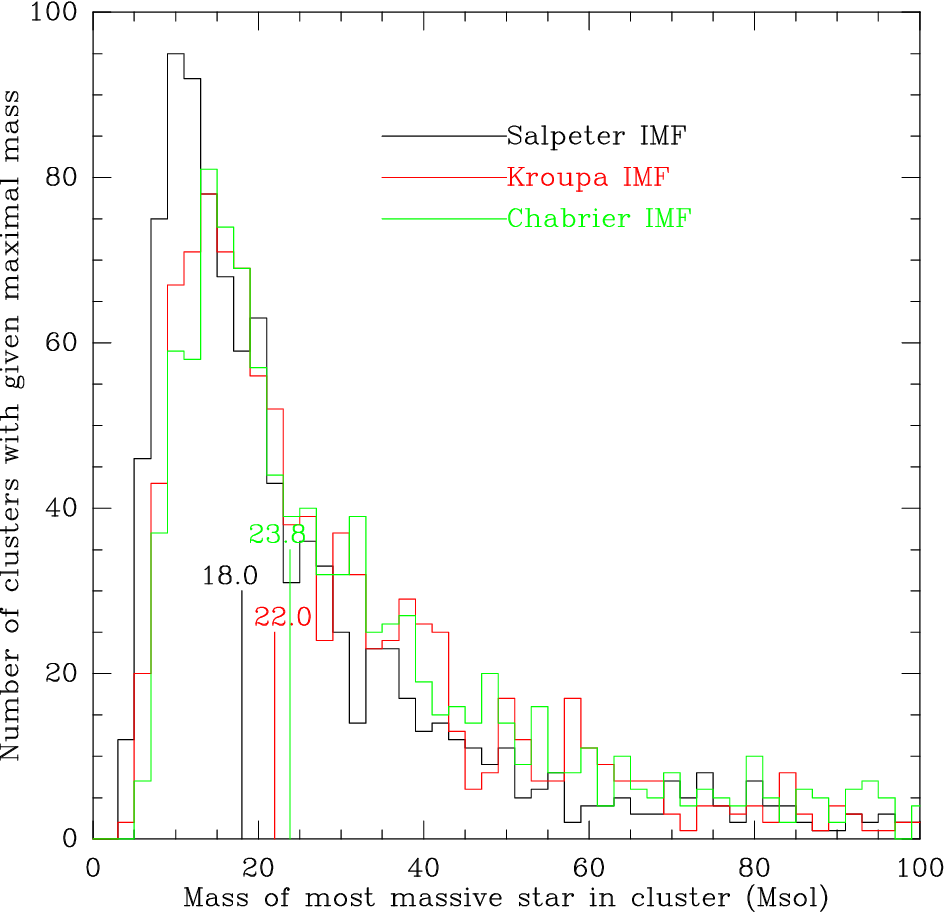}
        \includegraphics[width=\hsize{}]{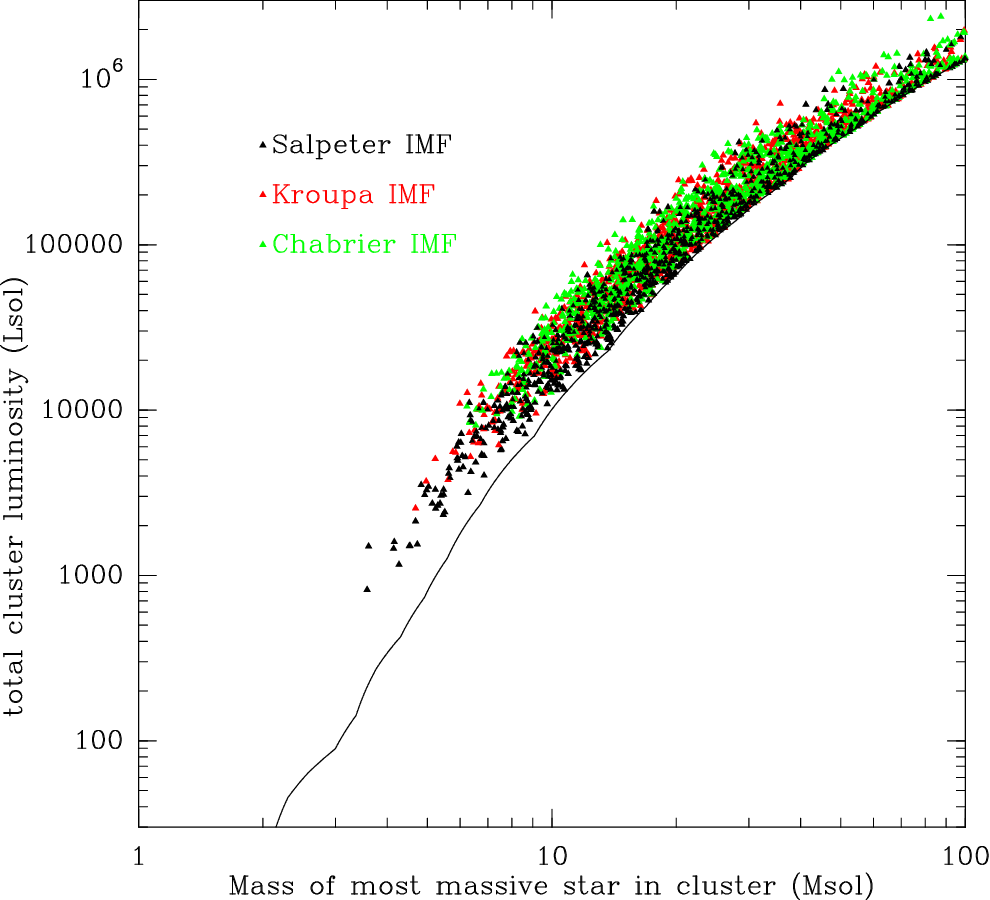}
        \caption{Even for a cluster with 300\msun\ of stars, the random sampling effect is huge.
        The top panel shows the distribution of the masses of the most massive star in each cluster, 
        with the median value indicated, for the 1000 simulated clusters.
        These median values are close to the maximal value suggested by \citet{Larson1982}.
        The bottom panel shows (1) that the luminosity of a 300\msun\ cluster can vary by a factor 1000 and (2) how important the most massive star is for the total zero-age luminosity.  The line indicates the luminosity of the single most massive star; when a very massive star is present, it dominates the luminosity (and ionizing photon production) of the cluster for its lifetime.}
        \label{randomness} 
\end{figure}

\begin{figure}
        \centering
        \includegraphics[width=\hsize{}]{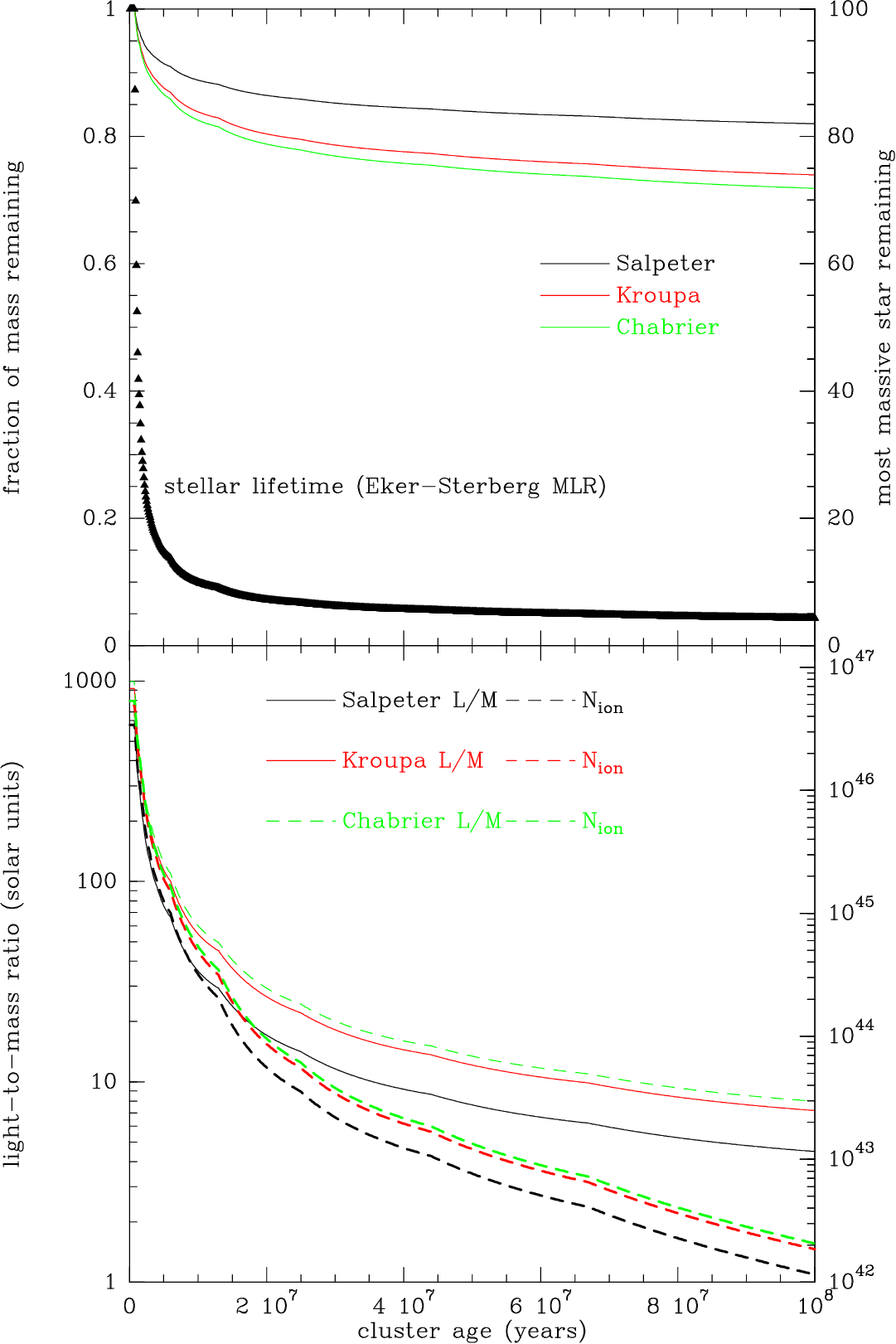}
        \caption{Time evolution of the stellar population in a cluster.  Black line is Salpeter, red is Kroupa, and green the Chabrier IMF.   
         The black triangles and dashed  lines show the property on the right-hand y-axis (Maximum stellar mass and number of ionizing photons per \msun ).}
        \label{starlife} 
\end{figure}
\end{appendix}

\end{document}